\PassOptionsToPackage{svgnames}{xcolor}
\documentclass[%
nofootinbib,
nobibnotes,
12pt
]{iopart}

\usepackage{hhline}
\usepackage{tikz}
\usetikzlibrary{plotmarks}

\usepackage{xcolor}

\usepackage{graphicx,caption,subcaption}
\usetikzlibrary{decorations.pathmorphing}

\usepackage{dcolumn}
\usepackage{bm}

\usepackage{hyperref}

\usepackage{mwe}

\usepackage{caption}
\usepackage{subcaption}

\renewcommand{\thesection}{}
\renewcommand{\thesubsection}{\arabic{section}.\arabic{subsection}}
\makeatletter
\def\@seccntformat#1{\csname #1ignore\expandafter\endcsname\csname the#1\endcsname\quad}
\let\sectionignore\@gobbletwo
\let\latex@numberline\numberline
\def\numberline#1{\if\relax#1\relax\else\latex@numberline{#1}\fi}
\makeatother

\usepackage{blindtext}
\usepackage [english]{babel}
\usepackage [autostyle, english = american]{csquotes}
\MakeOuterQuote{"}

\usepackage{titlesec}
\titleformat{\section}
  {\normalfont\sffamily\large\bfseries\color{black}}
  {\thesection}{1em}{}

\hypersetup{colorlinks=true,urlcolor=blue}

\renewcommand{\thesubsection}{\Alph{subsection}}

\usepackage{etoolbox}
\makeatletter
\def\@mkboth#1#2{}
\newlength\appendixwidth
\preto\appendix{\addtocontents{toc}{\protect\patchl@section}}
\newcommand{\patchl@section}{%
  \settowidth{\appendixwidth}{\textbf{Appendix }}%
  \addtolength{\appendixwidth}{1.5em}%
  \patchcmd{\l@section}{1.5em}{\appendixwidth}{}{\ddt}%
}
\makeatother

\usepackage{iopams}

\expandafter\let\csname equation*\endcsname\relax

\expandafter\let\csname endequation*\endcsname\relax

\usepackage{amsmath}

\usepackage{dcolumn}
\usepackage{epsf}
\usepackage{float}

\usepackage{tabularx}

\usepackage[final]{changes}

\usepackage{lipsum}
\definechangesauthor[color=blue]{.}
\colorlet{Changes@Color}{red}
\setremarkmarkup{(#2)}

\begin{document}

\title[]{\added{Universal Signatures of Majorana-like Quasiparticles in Strongly Correlated Landau-Fermi Liquids}}


\author{Joshuah T. Heath \& Kevin S. Bedell}

\address{Physics Department,  Boston  College,  Chestnut  Hill, Massachusetts  02467,  USA}
\ead{heathjo@bc.edu}
\vspace{10pt}
\begin{indented}
\item[]\today
\end{indented}

\begin{abstract} 
\added{Motivated by recent experiments in the Kitaev honeycomb lattice, Kondo insulators, and the "Luttinger's theorem-violating" Fermi liquid phase of the underdoped cuprates,} we extend the theoretical machinery of Landau-Fermi liquid theory to a \added{system of itinerant, interacting Majorana-like} particles. \added{Building upon a previously introduced model of "nearly self-conjugate" fermionic polarons,} a Landau-Majorana kinetic equation is introduced \added{to describe the collective modes and Fermi surface instabilities in a fluid of particles whose fermionic degrees of freedom obey the Majorana reality condition. At large screening, we show that the Landau-Majorana liquid harbors a Lifshitz transition for specific values of the driving frequency. Moreover, we find} the dispersion of the zero sound collective mode \added{in such a system, showing that there exists a specific limit where the Landau-Majorana liquid harbors a stability against Pomeranchuk deformations unseen in the conventional Landau-Fermi liquid. With these results, our work paves the way for possible extensions of the Landau quasiparticle paradigm to nontrivial metallic phases of matter.}
%
%
\end{abstract}

\vspace{2pc}

\submitto{\JPCM}

\maketitle

\tableofcontents

%


\section{I. Introduction}

Majorana fermions were originally introduced as neutral solutions to a symmetrized Dirac equation \cite{Majorana}. First proposed in the context of fundamental particle physics, Majorana particles have experienced a renaissance in the condensed matter community with the proposal that collective excitations in symmetry protected topological phases of matter \added{could} support self-conjugate edge excitations known as Majorana zero modes (MZMs)
\cite{Kitaev,FuKane,Yazdani,He,Bravyi,SuPeng,DasSarma,Freedman}.
More recently, quantum spin liquids (highly disordered spin systems with intrinsic topological order) have been shown to support Majorana excitations that are free to propagate through the bulk of the lattice \cite{Kitaev_honey}. Whereas the long-range entanglement of MZM pairs forbids the coherent definition of a Majorana number operator, the Majorana quasiparticle found in gapless spin liquids behaves more akin to a conventional complex fermion. The non-Abelian phase of the Kitaev spin liquid on the hyperoctagon lattice (a proposed model of $\beta$-Li$_2$IrO$_3$) is suggested to host a "Majorana metal" with a well-defined neutral Fermi surface \cite{Hermanns}, while recent Raman spectroscopy measurements on Majorana excitations in the Abelian phase of $\alpha$-RuCl$_3$ (a proposed realization of a 2D Kitaev honeycomb lattice) hints at low-temperature fermionic behavior \cite{Yiping}. 
%
%
%
%
%
%

The Majorana representation of some general spin-$1/2$ disordered state is not limited to the field of Kitaev materials. 
One of the first condensed matter applications of the Majorana representation beyond the simple spin-$1/2$ antiferromagnet\cite{Martin, Tsvelik, Shastry} was to a simple description of the low-energy spin Hamiltonian of the two-channel Kondo problem\cite{Coleman2,Coleman}. \added{Coleman et. al. performed this calculation by considering a simpler, "compactified" version of the two-channel problem, in which the local moment is coupled to the spin and isospin (i.e., charge) of a single-flavor liquid. Rewriting the spinor in terms of "scalar" and "vector" degrees of freedom, the Hamiltonian decouples, with that of the former defining itinerant Majorana fermions and that of the latter describing a local moment interacting with a sea of vector Majoranas. The presence of itinerant Majorana fermions in this mean-field description has led to the prediction that Kondo insulators have a band of neutral Majorana-like particles \cite{Coleman2,Coleman}, and has prompted an extensive experimental search for such particles in several candidate materials.}
The Kondo insulator SmB$_6$ has become of particular interest \added{in recent years}, as experiments appear to indicate the presence of a Fermi surface in the insulating phase of the material \cite{Tan,Sebastian}. Interestingly, ARPES measurements seem to suggest that such a conducting state is not the result of some topological properties of this specific Kondo insulator\cite{Hlawenka}, and has prompted the suggestion that SmB$_6$ (as well as certain Kitaev materials\cite{Baskaran2, Takikawa}) harbors a "Majorana-Fermi sea", where the Majorana reality condition imposes a severe retardation of hole-like excitations \cite{Baskaran, Fuhrman}. A "Landau-Majorana liquid", where the quasiparticles below the Fermi surface are severely supressed, would explain why Fermi liquid-like properties remain in the bulk of the low-temperature Kondo insulator\cite{Wakeham, Pachos}. \added{However, we cannot blindly associate the existence of a sharp discontinuity of the fermionic distribution with a finite weight of traditional Landau quasiparticles.} \added{Neutral Fermi surfaces of a similar nature are} seen in the $\mathbb{Z}_2$ fractionalized Fermi liquid theory \added{describing doped $d\ge 2$ Kondo lattice models \cite{Sachdev} and (possibly) the pseudogap phase of the cuprates \cite{Chatterjee} (assuming high magnetic field \cite{Proust})}, where 
the system simultaneously violates Luttinger's theorem. 
 \added{Nevertheless, all three materials discussed so far (the low-temperature regime of the Kitaev honeycomb candidates $\alpha$-RuCl$_3$ and lithium iridates; the bulk of the Kondo insulator SmB$_6$; and the high-field limit of certain underdoped cuprates) all share experimental evidence of a robust metallic state with neutral excitations in the absence of adiabatic continuity with the free Fermi-Dirac gas \cite{Grissonnanche,Yiping,Tafti,Tan,Sebastian}.}

In this paper, we propose a natural extension of Landau-Fermi liquid theory to \added{investigate and isolate the universal transport properties of a quantum fluid of self-conjugate Landau quasiparticles. In the topological materials usually considered to harbor Majorana-like excitations,  such particles are the direct result of either electron fractionalization or the formation of a bound state, both of which violate the underlying assumptions of Landau-Fermi liquid theory as there is no way to adiabatically connect the interacting eigenstates with those of the non-interacting Fermi gas. However, in a previous work} \cite{Heath}, we explicitly derived the low-temperature momentum distribution function of non-interacting particles \added{whose anti-symmetric degrees of freedom obey} the Majorana reality condition. \added{To maintain a non-zero chemical potential, microscopically we can consider these particles to be a gas of "Majorana polarons"}\footnote{\added{Note that, in \cite{Heath}, we called these Majorana-like particles "Majorana-Schwinger fermions", and their statistics "Majorana-Schwinger statistics". However, we have come to realize that such particles should be realized as more of a polaron like model, with the fermionic degrees of freedom obeying a self-conjugacy relation while remaining coupled to some residual bosonic degree of freedom, as opposed to a "purely" Majorana fermion description. We thank Prof. Natalia Perkins for making this suggestion.}}\added{; fermionic pairs asymmetrically dressed by bosonic excitations whose fermionic degrees of freedom mirror a non-interacting gas of self-conjugate particles. In this way, the total number of particles (fermions$+$bosons) is conserved, ensuring a well-defined number operator for the composite excitations.} We found that the mutual pairwise annihilation \added{of the Majorana polarons, while suppressed at zero temperature due to Pauli correlation, results in a sharp Fermi surface even in the presence of finite temperature. This latter fact motivates us to apply the techniques of Landau-Fermi liquid theory to the Majorana polaron gas in order to study an interacting system with self-conjugate Landau quasiparticles, as opposed to directly studying a system where purely fermionic degrees of freedom pair or fractionalize. In this way, the application of a Landau quasiparticle paradigm to a liquid of Majorana polarons allows us to take a step beyond the mean-field theory usually invoked when studying Majorana-like excitations, and permits us to explore the possible implications of self-conjugacy on select transport observables without having to take into consideration the topological non-trivialities inherent to other studies.
 }
%
%
%
%

The paper is structured as follows. \added{In II, we briefly review the non-interacting Majorana polaron model previously introduced in \cite{Heath}, and supply a generalization to allow for a smooth "transition" between the Landau-Majorana and Landau-Fermi liquid regimes. In III, we derive the Landau-Majorana kinetic equation, leading to a study of the distortions of the Majorana-Fermi sea in IV. In V, we derive explicit expressions for the zero sound collective mode of a Landau-Majorana liquid, ultimately cumulating to a discussion of Pomeranchuk instabilities and experimental predictions in VI. Conclusions follow in VII.} 

\section{II. Landau-like quasiparticles in a liquid of Majorana polarons} 

In order to build the theory of a Landau-Majorana liquid, we must first be able to coherently describe the eigenstates of the non-interacting system \cite{Landau1}. This was done in a previous work \cite{Heath}, where \added{the author of the present article considered the effects of self-conjugacy on traditional fermionic combinatorics. As mutual annihilation of neutral fermions cannot coexist with Pauli exclusion, such states must correspond to some dominant bosonic degree of freedom (DoF) in the system. By "tracing out" such bosonic terms, the minimization of the resulting thermodynamic potential allows us to derive a closed form for the distribution function of non-interacting particles with self-conjugate fermionic DoF. By taking the thermodynamic limit of the statistics,} the low-temperature momentum distribution $\widetilde{n}_{k\sigma}$ of the free Majorana-like gas is found to be
\begin{align}
\widetilde{n}_{k\sigma}^0&\approx \Theta(k_F-k)+\Theta(k-k_F)n_{k\sigma}^0
\end{align}
where $n_{k\sigma}^0$ is the momentum distribution of the free Fermi gas, and $\Theta(k_F-k)$, $\Theta(k-k_F)$ are Heaviside step functions. \added{These step functions are the direct result of mutual pairwise annihilation on the Majorana polaron's fermionic statistics, and may be interpreted as a suppressed quasiparticle excitation below the Fermi surface. We therefore see that, even at finite temperature, the regime $k<k_F$ is still characterized by a filled Fermi sea. This leads to exotic thermodynamic signatures of the Majorana polaron gas, including a modified Sommerfeld coefficient, a residual entropy near $T=0$, and a linear-$T$ dependence in the chemical potential. Microscopically, such a system can be realized in a modified Kitaev chain, where the usual Majorana operators are "dressed" with a bosonic pair, resulting in the usual Majorana-Kitaev Hamiltonian being mapped to a free Majorana polaron gas in the mean-field approximation \cite{Heath}.}


\added{In this work, we will take the preceding, non-interacting work a step further and consider a a Landau-Majorana liquid consisting of interacting Majorana polarons.} Even though the eigenstates of an Landau-Majorana liquid lack an isomorphism with the eigenstates of a non-interacting Fermi gas, we may still impose a one-to-one correspondence between the density fluctuations of the non-interacting and interacting Majorana systems. This is done by assuming, much as Landau did for the Landau-Fermi liquid, that the {bare} momentum distribution changes as an analytic function of the interaction strength. 
{This is similar to the Landau phenomenology considered in the ferromagnetic Fermi liquid, where an isomorphism is constructed between the eigenstates of the interacting spin excitation spectrum and the equilibrium magnetic system, as opposed to that of the traditional Fermi-Dirac distribution\cite{Dzy,Bedell2, Bedell3}.}
%
Although the self-conjugacy of the Majorana polarons leads to the possibility of mutual annihilation if we perturb the system, we assume the fraction of particles that annihilate is not dependent on the interaction strength. This is apparent from the low-$T$ description of the Kondo insulator SmB$_6$, as a robust Fermi surface (the hallmark of the Majorana-Schwiner gas) and a Landau-Fermi liquid-like linear-$T$ specific heat is found in the bulk even in the presence of strong interactions \cite{Wakeham,Sebastian}.

\begin{figure}[htbp]
\hspace*{-10mm} 
\begin{subfigure}{.55\columnwidth}
\includegraphics[width=1\columnwidth]{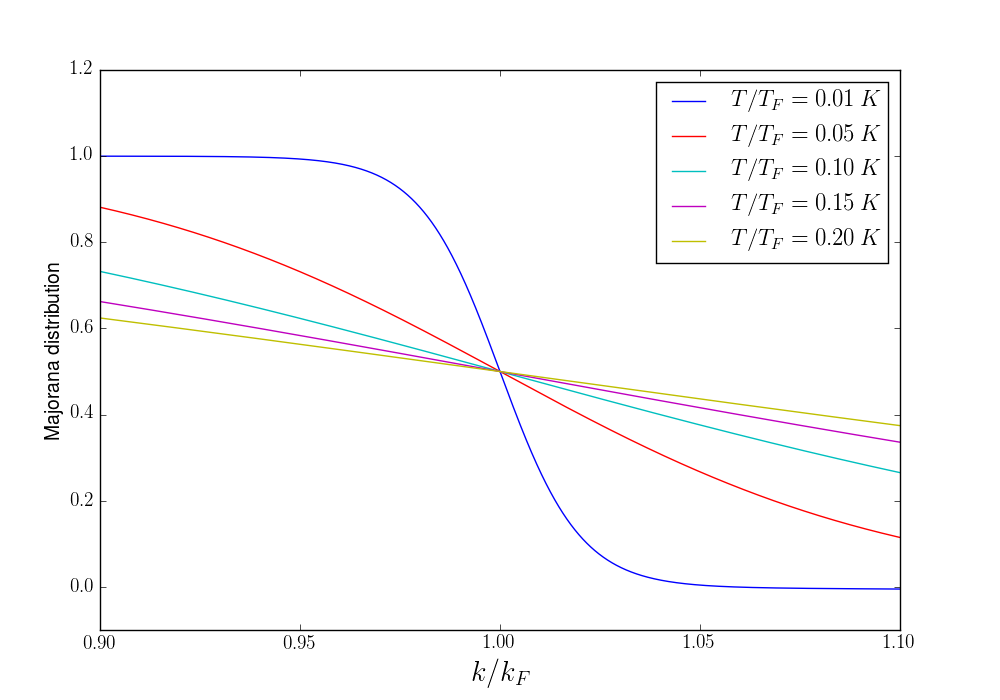}
  \caption{$\alpha/k_F>1.0$}\label{fig:2a} 
\end{subfigure}
\begin{subfigure}{.55\columnwidth}
\includegraphics[width=1\columnwidth]{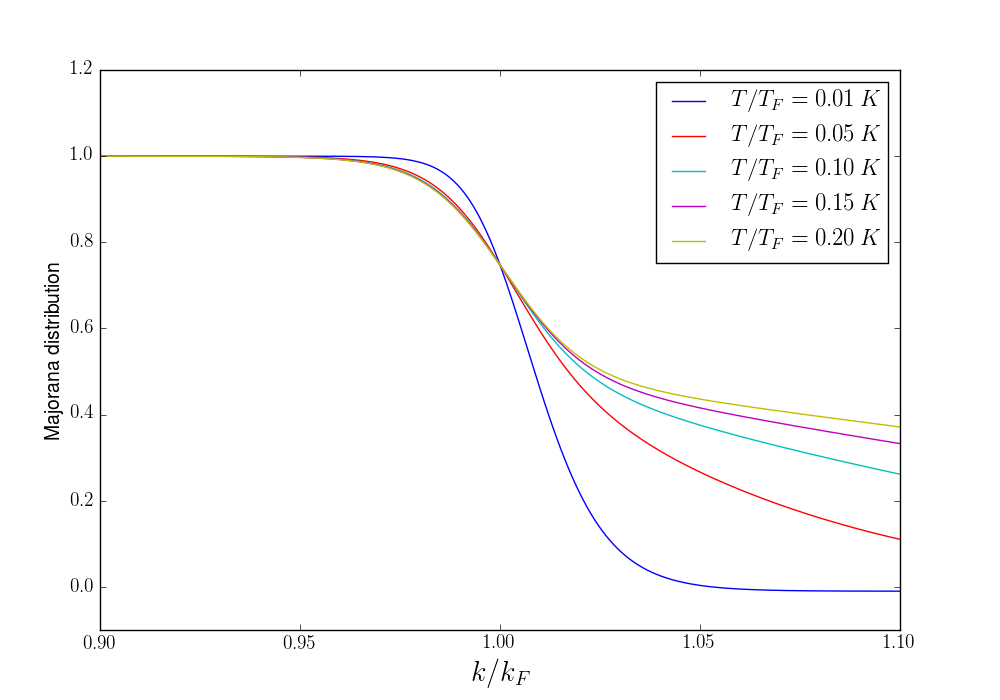} 
  \caption{$\alpha/k_F=0.01$}\label{fig:2b} 
\end{subfigure}
\hspace*{-10mm} 
\begin{subfigure}{.55\columnwidth}
 \includegraphics[width=1\columnwidth]{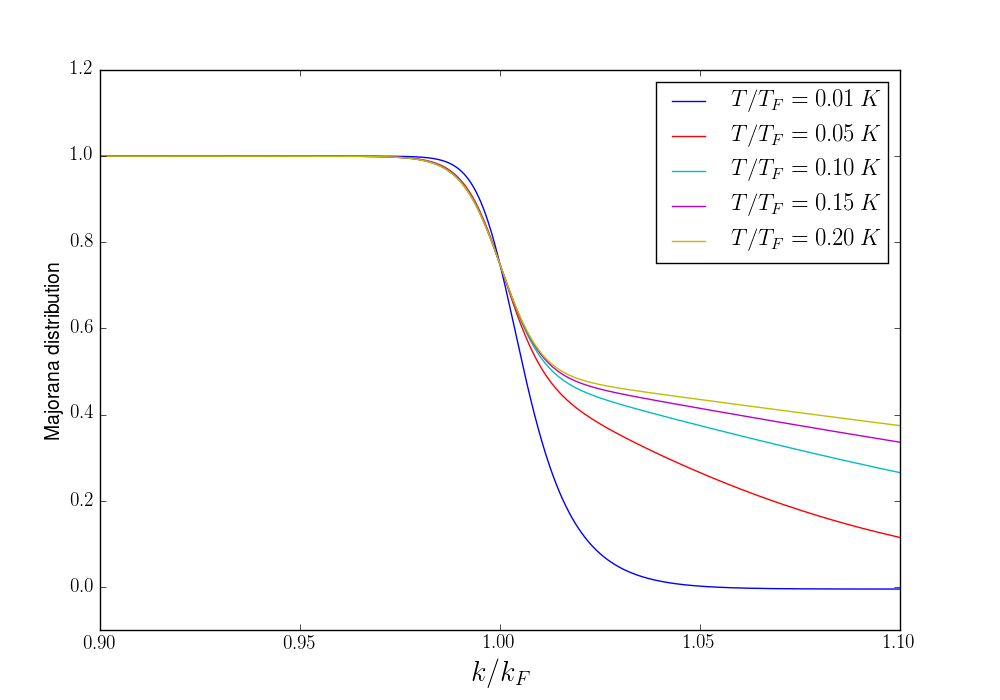} 
   \caption{$\alpha/k_F=0.005$}\label{fig:2c} 
\end{subfigure}
\begin{subfigure}{.55\columnwidth}
\includegraphics[width=1\columnwidth]{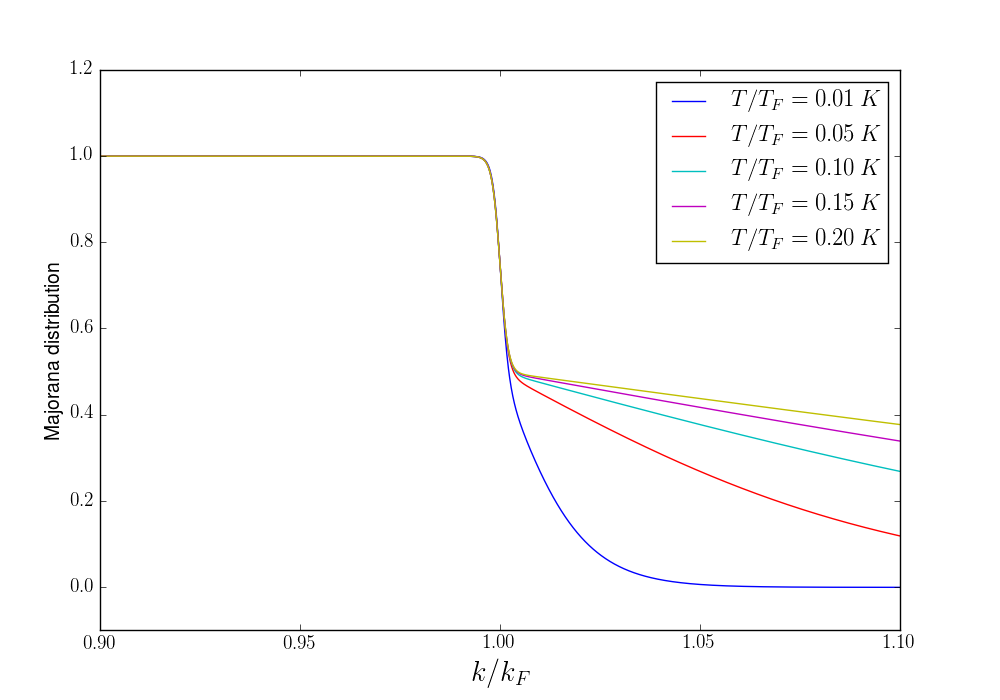} 
   \caption{$\alpha/k_F=0.001$}\label{fig:2d} 
\end{subfigure}
\caption{{\color{black} 
\added{The Majorana-Fermi distribution for several values of $\alpha/k_F$. The Majorana parameter $\alpha$ is an experimentally-determined tool that allows us to tell how close our physical system is to the limit of a "true" Majorana polaron gas, whichcorresponds to $\alpha/k_F\rightarrow 0$. For large $\alpha/k_F$, the Majorana-Fermi distribution is identical to the Fermi-Dirac distribution. However, as $\alpha/k_F$ gets closer to zero, a robust Fermi surface emerges at finite temperature (as derived explicitly in the statistical model introduced in \cite{Heath}).}
}} 
\label{Fig1}
\end{figure}

The above argument allows us to write the density fluctuation $\delta \widetilde{n}_{k\sigma}$ in the Landau-Majorana liquid as
\begin{align}
\widetilde{n}_{k\sigma}-\widetilde{n}_{k\sigma}^0\equiv \delta \widetilde{n}_{k\sigma}=\Theta (k-k_F)\delta n_{k\sigma}
\end{align}
We see that, as in previous descriptions of the Majorana-Fermi sea\cite{Baskaran}, the quasihole state is significantly suppressed. \added{Nevertheless, we must remember that the Heaviside step functions present in the Majorana distribution are approximations only valid in the thermodynamic limit. More concisely, the Majorana distribution corresponds to a fermionic distribution with severely suppressed "smearing" for $k<k_F$, and not necessarily defined by an absolute discontinuity at the Fermi momentum.} Therefore, because the Heaviside step function is an approximation only valid in the thermodynamic limit \cite{Heath}, we will rewrite it here as a Fermi-like function near the Majorana-Fermi surface:
\begin{align}
\Theta(k-k_F)\approx \frac{2-e^{-\frac{\alpha}{k_F}}}{1+e^{-\frac{k-k_F}{\alpha}}}\equiv \left(2-e^{-\frac{\alpha}{k_F}}\right)\mathfrak{F}(k,\,\alpha) \label{eq3}
\end{align}

\noindent where $\alpha$ is a tunable parameter that we call the Majorana constant or parameter. To ensure the exponential remains unitless, we take $\alpha$ to have the same units as $k_F$. Such a constant parametrizes the finite-system effects not taken into account in the derivation of the Majorana polaron's Boltzmann entropy. \added{The exact value of $\alpha$ would, in general, be system dependent and would have to be determined experimentally. However, it is clear to see that the limit $\alpha\rightarrow 0$ corresponds to the "sharp" discontinuity expected in the thermodynamic limit of a Majorana polaron gas, while $\alpha\rightarrow \infty$ allows us to recover regular Fermi-Dirac-like behavior. The additional multiplicative term $2-e^{-\frac{\alpha}{k_F}}$ is included to ensure that we have a smooth transition to complex fermionic behavior in the large-$\alpha$ limit (i.e., $\alpha/k_F>>1$). This term wasn't included in the system discussed in \cite{Heath}, as in such a paper we were not concerned about a "smooth" transition to a Landau-Fermi liquid state as we are here. In a similar fashion, we will take the convention that, in the limit of $\alpha\rightarrow 0$ and $k=k_F$ in the above discussion, that $\mathfrak{F}(k,\,\alpha)\approx 1$. This was of little concern in \cite{Heath}, as the main concern in the study of the non-interacting Majorana gas was the thermodynamics as opposed to the nature of excitations in the vicinity of the Fermi surface. In this paper, we more carefully define this function in close proximity to $k=k_F$ in preparation for the study of a Landau-Majorana liquid.} 


\added{It is important to note that the thermodynamic signatures that follow from the Majorana polaron distribution defined in \cite{Heath} is explored for d=1, 2, and 3 dimensions. However, in the following discussion, we will consider an interacting Majorana polaron system in terms of a Landau-like quasiparticle, and thus we will restrict our discussion to dimension $d\ge 2$. This is because a Landau-like quasiparticle breaks down in $d=1$ due to the lack of a coherent particle-hole excitation. This leads to the existence of a Tomonaga-Luttinger liquid as opposed to a Landau-Fermi liquid, and it is for this reason that we restrict ourselves to two dimensions or higher.}
%

\section{III. Majorana-Landau Kinetic Equation}

\added{For a regular Landau-Fermi liquid, the starting point is often an expansion of the "excitation" free energy $F-F_0$ in terms of the difference $\delta n_k$ of the interacting and non-interacting distributions:}
\begin{align}
\added{F-F_0=\sum_k(\epsilon_k-\mu)\delta n_k+\frac{1}{2}\sum_{kk'}f_{kk'}\delta n_k \delta n_{k'}}
\end{align}
\added{In the above, the coefficient $f_{kk'}$ is known as the Landau parameter, defined as the second variational derivative of the free energy with respect to $n_k$. Such a parameter quantifies the interaction of the quantum fluid.}

The functional expansion of the Landau-Majorana free energy is analogous to the Landau-Fermi case \added{given above, except that the departure $\delta n_k$ from the non-interacting distribution is replaced by $\delta \widetilde{n}_k$; i.e., that the interacting system experiences a severe particle-hole asymmetry by virtue of the underlying quantum statistics as opposed to some strong interaction. The resultant excitation free energy for the interacting Majorana polaron system can} be written in two equivalent forms:
\begin{align}
F-F_0&=\sum_k (\epsilon_k-\mu)\delta \widetilde{n}_k+\frac{1}{2}\sum_{kk'}f_{kk'}\delta \widetilde{n}_k\delta \widetilde{n}_{k'} \label{eq4}\\
&=\sum_k(\widetilde{\epsilon}_k-\mu\mathfrak{F}(k,\,\alpha))\delta n_k+\frac{1}{2}\sum_{kk'}\widetilde{f}_{kk'}\delta n_{k'}\delta n_{k'}\label{eq5}
\end{align}
where $\widetilde{\epsilon}_k$ and $\widetilde{f}_{kk'}$ are the regular Landau-Fermi liquid quasiparticle energies and interaction parameters multiplied by $\mathfrak{F}(k,\,\alpha)$ and $\mathfrak{F}(k,\,\alpha)\mathfrak{F}(k',\,\alpha)$, respectively. 
%
%
%
Eq. \eqref{eq4} is interpreted as a suppression of quasihole excitations while maintaining Fermi liquid-like interactions. In contrast, Eq. \eqref{eq5} may be interpreted as a system with Fermi liquid-like excitations with a suppressed interaction for $k<k_F$. We can therefore regard the Landau-Majorana liquid as a regular Landau-Fermi liquid with either retarded quasihole excitations or with a suppressed interaction term below the Majorana-Fermi surface. This allows us to define the Landau-Majorana effective mass $\widetilde{m}^*$ in terms of the Landau-Fermi effective mass $m^*$ (see Appendix A), which gives us $\widetilde{m}^*/m^*\approx k_F/8\alpha$. As such,
the Landau-Majorana liquid may likewise be considered to be a Landau-Fermi liquid with an effective mass rescaled by $\alpha$, and subsequently a severely suppressed particle-hole continuum, given by
$
(\widetilde{\epsilon}_{k+q}-\widetilde{\epsilon}_k)/\epsilon_{k_F}=\frac{8\alpha}{k_F^3}\left(q^2+2k\cdot q\right)
$. A large effective mass is similarly seen from quantum oscillation measurements in the Kondo insulators YbB$_{12}$\cite{Liu} and the ($011$)-plane of SmB$_6$\cite{Luo}, which have already been suggested to harbor a neutral Majorana metallic state \cite{Coleman2, Tan, Baskaran, Fuhrman}.

\added{With the non-interacting Majorana polaron distribution and its interacting limit appropriately defined, we can now move onto the implications of the interacting theory. Because we propose that the Landau quasiparticle picture is still viable in our Majorana polaron system due to the existence of a robust Fermi surface, we expect the thermodynamic hallmarks of the non-interacting Majorana polaron gas discussed in \cite{Heath} to similarly describe the basic thermodynamic behavior seen in an interacting Landau-Majorana liquid, albeit with a slightly different effective mass. We will therefore dedicate the rest of this paper to universal signatures of the Landau-Majorana system inherent to the interacting system; i.e., collective excitations and Fermi surface instabilities.}

Taking the above \added{approach}, we are now in a position to write down the collision integral for the Landau-Majorana liquid. In terms of the Majorana polaron distribution function, the collision integral is given by

\begin{align}
I(\widetilde{n}_k)
&=\frac{d}{dt}\widetilde{n}_k=\frac{\partial}{\partial t} \widetilde{n}_k-\left\{\epsilon_k,\,\widetilde{n}_k\right\}_{PB}
\end{align}
Because we restrict ourselves to quasiparticle states just above the Majorana-Fermi surface, the Landau-Majorana dispersion relation becomes $\widetilde{n}_k\approx\Theta_{\gtrapprox}(k-k_F)n_k$, where $n_k$ is the Landau-Fermi liquid dispersion, and the Poisson bracket is given by 
\begin{align}
\left\{\epsilon_k,\, \widetilde{n}_k\right\}_{PB}&=\nabla_r\epsilon_k \cdot \nabla_k\widetilde{n}_k-\nabla_k\epsilon_k \cdot \nabla_r \widetilde{n}_k\notag\\
&=\Theta(k-k_F)\nabla_r \epsilon_k\cdot \nabla_k n_k+n_k\nabla_r \epsilon_k \cdot \nabla_k \Theta(k-k_F)-\nabla_k \epsilon_k\cdot \nabla_r (\Theta(k-k_F) n_k)\notag\\
&\approx \Theta(k-k_F){\bf q} \cdot {\bf v}_k\frac{\partial n_k^0}{\partial \epsilon_k}\delta \epsilon_k+n_k^0  {\bf q}\cdot \nabla_k \Theta(k-k_F)\delta \epsilon_k-\Theta(k-k_F){\bf q}\cdot {\bf v}_k \delta n_k
\end{align}

\added{It is important to note that the second term in the above is unique to the Landau-Majorana system, 
as it originates from the inherent hole suppression in the Majorana polaron statistics. In the non-interacting case, we have quasi-hole suppression isotropically across the entire Fermi surface at finite temperature by virtue of the fermionic self-conjugacy, as the smearing from thermal excitations is isotropic (i.e., there is no thermal gradient over $k=k_F$). In the interacting system we consider in this article, the "smearing" is the direct result of a perturbation along the $\hat{q}$-direction; i.e., the "sharp" features of the distribution function remain in the direction of particle-hole propagation only at $T=0$. As such, we can write this term as }
\begin{align}
\added{n_k^0  {\bf q}\cdot \nabla_k \Theta(k-k_F)\delta\epsilon_k= n_k^0 q\delta(k-k_F)\delta\epsilon_k}
\end{align}
\added{This divergent term is the direct result of an exceptionally "sharp" point in the distribution along the direction of particle-hole propagation only, and thus will be independent of the relative angle between ${\bf q}$ and ${\bf k}$.} 

Returning to the form of the Landau-Majorana collision integral, we can now write 
\begin{align} 
iI( \widetilde{n}_k)
&=\Theta(k-k_F)I(n_k)-n_k^0 q \delta (k-k_F)\delta\epsilon_k
\end{align}

\noindent where $I(n_k)$ is the Landau-Fermi collision integral. To simplify the above, we can make an approximation of the Heaviside theta function given in Eq. \eqref{eq3}. Hence,

\begin{align}
\delta(k-k_F)&=\frac{\partial}{\partial k}\Theta(k-k_F)\notag\\
&\approx (2-e^{-\frac{\alpha}{k_F}})\frac{\partial}{\partial k}\mathfrak{F}(k,\,\alpha)\notag\\
&=\frac{2-e^{-\frac{\alpha}{k_F}}}{\alpha}\mathfrak{F}(k,\,\alpha)\left(1-\mathfrak{F}(k,\,\alpha)\right)
\end{align}
Writing out the full form of $I(n_k)$, the Landau kinetic equation then becomes
\begin{align}
iI( \widetilde{n}_k)=\left(2-e^{-\frac{\alpha}{k_F}}\right)\mathfrak{F}(k,\,\alpha)\left\{\delta n_k \bigg(\omega-{\bf q}\cdot {\bf v}_k \bigg)+ \left({\bf q}\cdot {\bf v}_k\frac{\partial n_k^0}{\partial \epsilon_k}+\frac{q}{\alpha}\left\{1-\mathfrak{F}(k,\,\alpha)\right\}n_k^0\right)\delta \epsilon_k\right\}\label{eq8}
\end{align}

\noindent In the limit of $T\rightarrow 0$, we asymptotically approach the collisionless regime. The term in the brackets is equal to zero, and we are left with the Landau-Majorana (LM) kinetic equation\cite{Silin}. \added{Note that the term proportional to $\frac{q}{\alpha}$ is unique to the Landau-Majorana system, and will lead to transport properties unseen in the regular, garden-variety Landau-Fermi liquid when $\alpha<q$. Interestingly, in the limit of $\alpha>>q$, the above reduces to the regular Landau kinetic equation defining excitations in a Landau-Fermi liquid. This behavior of the Majorana parameter (i.e., the fact that we recover Landau-Fermi liquid behavior in the limit of $\alpha\rightarrow \infty$) will be seen in future calculations.}
\section{IV. Distortions of the Majorana-Fermi sea} \added{We now want to understand how the Majorana-Fermi surface reacts to an external perturbation.} We may do this by expressing Eqn. \eqref{eq8} in terms of the dimensionless parameters $s=\omega/qv_F$ and $\cos\theta=\hat{\bf q}\cdot \hat{\bf v}_k$. First, we write the differential element $\delta \epsilon_k$ as
\begin{align}
\delta \epsilon_k&=U+\int \frac{d\Omega_{k'}}{4\pi} f_{kk'}\delta n_{k'}\notag\\
&\equiv U+\int_{k'} f_{kk'}\delta n_{k'}
\end{align}
Taking the limit of $U\rightarrow 0$ and using the fact that the Fermi surface distortion \added{$\nu_{k'}$} is given by
\begin{align}
\delta n_{k'}=-\frac{\partial n_{k'}^0}{\partial \epsilon_{k'}}\nu_{k'}
\end{align}
We obtain
\begin{align}
\nu_k &+\frac{\cos\theta}{\cos\theta-s}\int_{k} f_{kk'}\frac{\partial n_{k'}^0}{\partial \epsilon_{k'}}\nu_{k'}+\frac{1}{\cos\theta-s}\left\{\frac{1-\mathfrak{F}(k,\,\alpha)}{\alpha v_F}\right\}\frac{n_k^0}{\partial n_k^0/\partial \epsilon_k}\int_{k'} f_{kk'}\frac{\partial n_{k'}^0}{\partial \epsilon_{k'}}\nu_{k'}=0
\end{align}
\added{As stated before, a non-zero value of the final term proportional to $1/\alpha$ is specific to the Landau-Majorana liquid, with the limit $\alpha\rightarrow \infty$ resulting in the traditional Landau-Fermi liquid equation.}
\added{We can simplify it by noting that \cite{Bedell}}
\begin{align}
\frac{\delta n_{k'}}{\delta n_k}=\frac{q\cdot v_{k'}}{\omega-q\cdot v_{k'}}\frac{\partial n_{k'}^0}{\partial \epsilon_{k'}}\frac{\delta \epsilon_{k'}}{\delta n_{k'}}
\end{align}
The value of the left-hand term differs depending on whether we are dealing with the unscreened or screened limits, defined by $\lim_{\substack{q\rightarrow 0 \\ \omega\rightarrow 0 }}\frac{\delta n_{k'}}{\delta n_k}=0$ and $\lim_{\substack{w\rightarrow 0 \\ q\rightarrow 0 }}\frac{\delta n_{k'}}{\delta n_k}=\frac{\partial n_{k'}^0}{\partial \epsilon_{k'}}a_{kk'}^s$ respectively, where $a_{kk'}^s$ is the quasiparticle scattering amplitude \cite{Baym}. These two limits result in the above being recast in the form
\begin{align}
\frac{\partial n_{k'}^0/\partial \epsilon_{k'}}{\partial n_k^0/\partial \epsilon_k}&=\frac{\delta n_{k'}}{\delta n_k}\frac{\nu_k}{\nu_{k'}}\rightarrow \begin{cases}
\frac{\partial n_{k'}^0}{\partial \epsilon_{k'}}\frac{\nu_k}{\nu_{k'}}a_{kk'}^s,\quad &\omega \rightarrow 0,\quad q\rightarrow 0 \notag \\
0,\quad &q \rightarrow 0,\quad \omega\rightarrow 0
\end{cases}
\end{align}
Where the quasiparticle scattering amplitude is given by
\begin{align}
a_{kk'}^s&\equiv \lim_{\substack{\omega\rightarrow 0 \\ q\rightarrow 0}}\frac{\delta \epsilon_{k'}}{\delta n_{k'}}
\end{align}


\noindent This simplification leads us to recast the Landau kinetic equation terms of the Majorana-Fermi surface distortion $\nu_k$: 

%
%
\begin{align}
\nu_k\left\{
1+\frac{\mathbb{F}(\omega,\,q)}{4\alpha v_F }\left(\frac{1}{\cos\theta-s}\right)
\right\}+\frac{\cos\theta}{\cos\theta-s}\int_{k'}f_{kk'}\frac{\partial n_{k'}^0}{\partial \epsilon_{k'}^0}\nu_{k'}=0\label{eq11}
\end{align}
\\
\noindent where we have taken $n_k^0(1-\mathfrak{F}(k,\,\alpha))\approx \frac{1}{4}$ for $k\gtrapprox k_F$, and 
\begin{align}
\mathbb{F}(\omega,\,q)=\begin{cases}
\int_{k'} f_{kk'}\frac{\partial n_{k'}^0}{\partial \epsilon_{k'}}a_{kk'}^s,\quad &\textrm{screened}\\
\mathcal{C},\quad &\textrm{unscreened}
\end{cases}
\end{align}
where $\mathcal{C}$ is some constant \added{with units of energy. The numerical value of $1/4$ is the direct result is the direct result of taking $k \approx k_F$, where $n_k^0\approx \frac{1}{2}$ and $\mathcal{F}(k,\,\alpha)\approx \frac{1}{2}$ in the limit of $\alpha\rightarrow 0$. For the limit of $\alpha\rightarrow \infty$ we will consider shortly, this value of $1/4$ will instead be $1/2$ for reasons considered in the context of the Heaviside function above, but as all such terms go as $\frac{1}{\alpha}$ in such a limit, the "Majorana-specific" terms disappear and we may ignore such factors of two.}

\added{ In the above, we can see that the} term proportional to $\mathbb{F}(\omega,\,q)$ is unique to the interacting Majorana system. { Also note that in all calculations concerning zero sound in this paper, we assume the main interesting features in the collective excitations are from the $\ell=0$ channel. We subsequently truncate all spherical harmonics to this order.}
%
%
\added{
For the unscreened Landau-Majorana liquid, the non-interacting limit $f_{kk'}\rightarrow 0$ yields the following value of the constant $\mathcal{C}$:}

\begin{align}
\mathcal{C}=4\alpha v_F (s-\cos\theta)\lim_{f_{kk'}\rightarrow 0}\nu_k
\end{align}

%
%
\added{\noindent Because the definition of $\nu_k$ is the energy by which zero sound splits the quasiparticle distribution\cite{Baym}, $\lim_{f_{kk'}\rightarrow 0}\nu_k=0$, and hence $\mathcal{C}=0$. This tells us that the kinetic equation of the unscreened Landau-Majorana liquid is identical to that of the traditional Landau-Fermi liquid:}
\begin{align}
\nu_k&=\frac{\cos\theta}{s-\cos\theta}\int_{k'} f_{kk'}\frac{\partial n_{k'}^0}{\partial \epsilon_{k'}^0}\nu_{k'}\label{unscreened}
\end{align}

\noindent \added{The above tells us that, in the limit of unscreened interactions, the effective excitations of the Landau-Majorana liquid are identical to those in the Landau-Fermi liquid. This makes sense, as the screening length of some interaction should go as the inverse density (such as in Debye-H\"uckel). As the screening length becomes large comparable to $\alpha$, the Majorana polaron gas becomes dilute, and collisions between particles become less of a concern. Hence, in such a limit, the collective excitations of a Landau-Majorana liquid should reduce to that of the Landau-Fermi liquid, as seen in Eqn. \eqref{unscreened}.
}

\added{Compare the above result to the corresponding equation for the screened Landau-Majorana liquid, where the equation for $\nu_k$ can be simplified to the form}

%
%
%
%
%
%

\begin{align}
\nu_k&=\frac{\cos\theta}{s-\cos\theta} \left(
\frac{1}{
1+\frac{1}{\gamma(\cos\theta-s)}}
\right)\int_{k'} f_{kk'}\frac{\partial n_{k'}^0}{\partial \epsilon_{k'}^0}\nu_{k'}\label{screened}
\end{align}
where
\begin{align}
\gamma =4\alpha v_F\left(\int_{k'} f_{kk'}\frac{\partial n_{k'}^0}{\partial \epsilon_{k'}^0}a_{kk'}^s\right)^{-1}\label{eq15}
\end{align}



\noindent The limit of $\gamma\rightarrow 0$ is interpreted as a maximized "Majorana-like" contribution to the momentum distribution function, and therefore a total suppression of quasihole excitations. For no interaction, the system will be "purely" Majorana-like. For a very strong repulsive interaction, the system will be equivalent to that of a Fermi-Dirac system or nearly so due to a suppressed annihilation cross-section.

\begin{figure*}[htbp]
\hspace*{-10mm} 
\begin{subfigure}{.55\columnwidth}
\includegraphics[width=1\columnwidth]{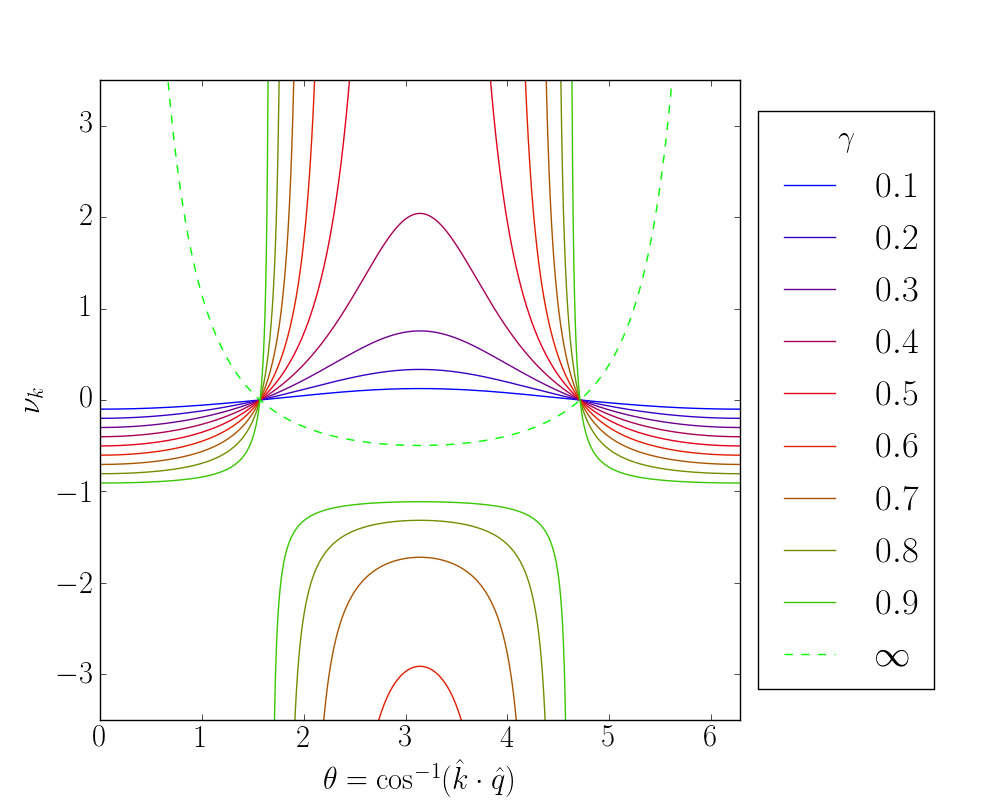}
  \caption{$s\sim 1.0$}\label{fig:2a} 
\end{subfigure}
\begin{subfigure}{.55\columnwidth}
\includegraphics[width=1\columnwidth]{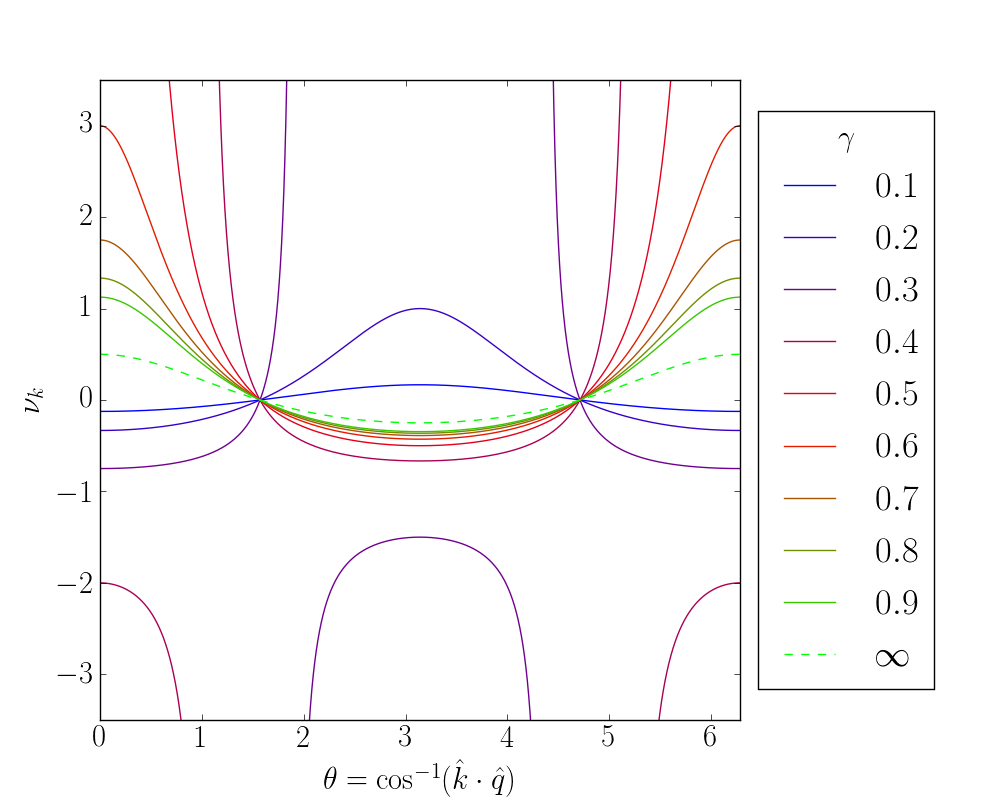} 
  \caption{$s=3.0$}\label{fig:2b} 
\end{subfigure}
\hspace*{-10mm} 
\begin{subfigure}{.55\columnwidth}
 \includegraphics[width=1\columnwidth]{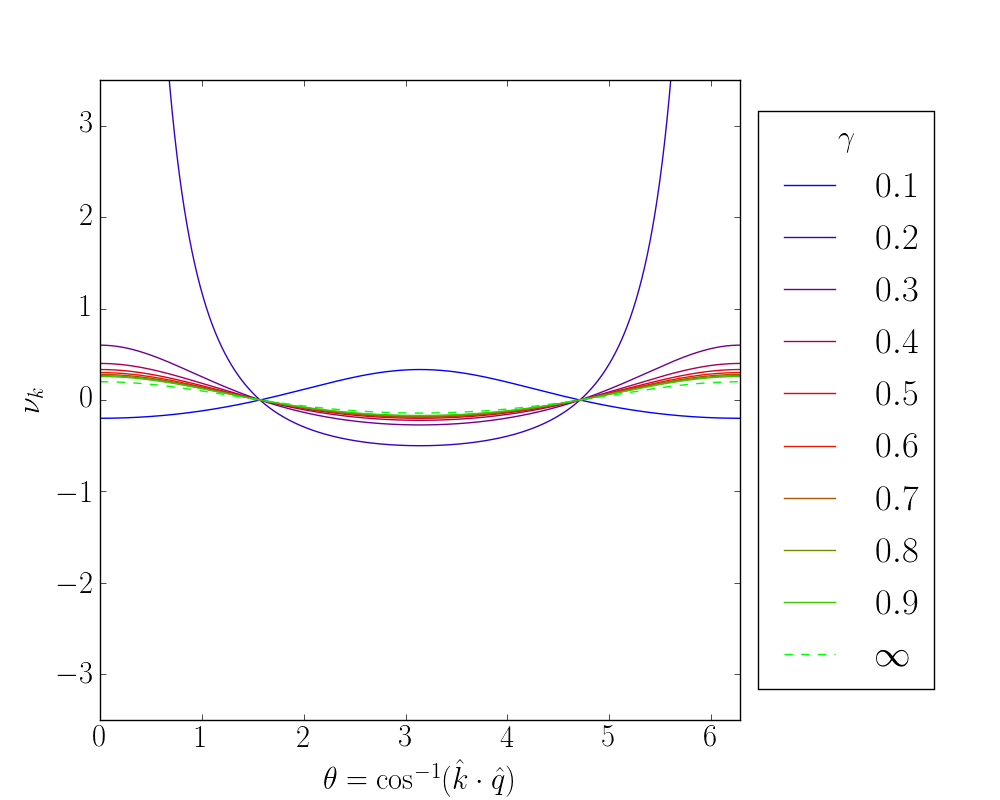} 
   \caption{$s=6.0$}\label{fig:2c} 
\end{subfigure}
\begin{subfigure}{.55\columnwidth}
\includegraphics[width=1\columnwidth]{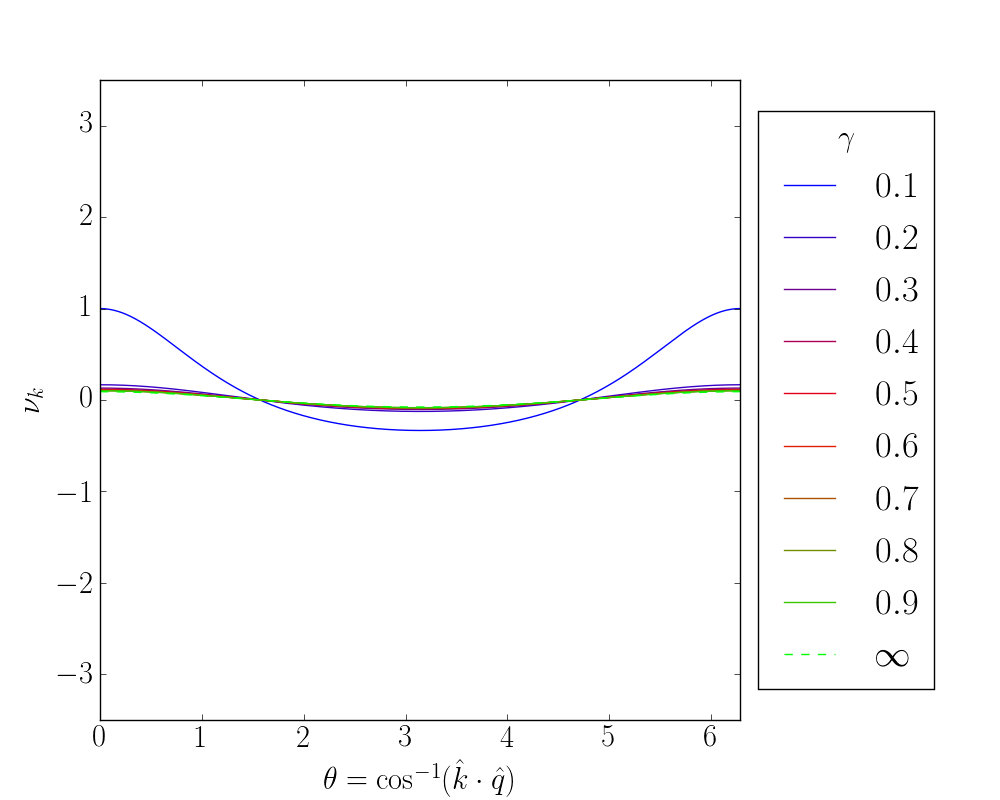} 
   \caption{$s=12.0$}\label{fig:2d} 
\end{subfigure}
\caption{{\color{black} 
Fermi surface distortion for the Landau-Fermi liquid (dotted line) and Landau-Majorana liquid (solid lines) vs. $\theta$ for \added{several} fixed values of $s=\omega/q v_F$ \added{across several values of $\gamma$}. \added{As we increase the value of $\gamma$ (i.e., approach the Landau-Fermi liquid of $\alpha\rightarrow \infty$), a Lifshitz transition occurs which heralds a change in the fundamental topology of the Fermi surface. As $s$ increases, the critical value of $\gamma$ at which this Lifshitz transition occurs decreases. This tells us that a Landau-Fermi liquid extension of a Majorana fermion-like model with conserved total particle number (i.e., the "Majorana polaron") experiences a Fermi surface instability for large Landau parameters beyond the usual Pomeranchuk instability.  }
}} 
\label{Fig2}
\end{figure*}

%
%
%
%

The collective $\ell=0$ breathing mode (zero sound\cite{Landau2,Mermin}) for the Landau-Fermi ($\gamma\rightarrow \infty$) result is apparent in the above equations, and is the signature of dominant forward-scattering between quasiparticles. For the Landau-Majorana liquid, the "purely" \added{Majorana-like} Majorana-Fermi surface retains its isotropic shape while simultaneously being shifted by some small amount. Such behavior is reminiscent of the $\ell=1$ breathing mode (first sound) observed in the Landau-Fermi liquid, except with a net "backscattering" (i.e., $\theta=\pi$) contribution. Emergent backscattering is evident from the unusually large effective mass $\widetilde{m}^*$ in the Landau-Majorana liquid, which subsequently leads to an increased number of interactions with large momentum transfer.

{ In Fig. \ref{fig:Fig2}, we see the Fermi and Majorana-Fermi surface distortions for fixed values of the collective mode velocity $s=\omega/qv_F$ plotted vs. various values of $\gamma$. }
%
%
%
%
%
%
%
%
%
As the value of $\gamma$ passes through a certain threshold, the Majorana-Fermi surface experiences exponential divergence until it eventually settles into the regular zero sound behavior of an Landau-Fermi liquid. From Eq. \eqref{eq15}, singularities occur when
$
\hat{k}\cdot \hat{q}=\cos^{-1}\left(s-1/\gamma\right)
$.
As $\gamma\rightarrow \infty$, we recover the standard Landau damping of a Landau-Fermi liquid\cite{Landau3,Pines}.

We interpret this singularity as the onset of a change of the Fermi surface topology brought about by increasing $\gamma$ and, hence, the gradual screening of the effects of self-conjugacy in the quantum fluid.  The exact location of this change in topology can be solved for exactly, yielding
\begin{align}
\cos\theta =s-\frac{1}{4\alpha v_F}\int_{k'} f_{kk'}\frac{\partial n_{k'}^0}{\partial \epsilon_{k'}^0}a_{kk'}^s
\end{align}
This allows us to reduce the Landau kinetic equation at the Lifshitz transition to
\begin{align}
\frac{\frac{1}{4\alpha v_F}\int_{k'} f_{kk'}\frac{\partial n_{k'}^0}{\partial \epsilon_{k'}^0}a_{kk'}^s-s}{\frac{1}{4\alpha v_F}\int_{k'} f_{kk'}\frac{\partial n_{k'}^0}{\partial \epsilon_{k'}^0}a_{kk'}^s}\int_{k'}f_{kk'}\frac{\partial n_{k'}^0}{\partial \epsilon_{k'}^0}\nu_{k'}=0
\end{align}
There are two possible solutions to the above equation. Either we can have

\begin{align}
s=\frac{1}{4\alpha v_F}\int_{k'} f_{kk'}\frac{\partial n_{k'}^0}{\partial \epsilon_{k'}^0}a_{kk'}^s
\end{align}

\noindent Or we can have 

\begin{align}
\int_{k'} f_{kk'}\frac{\partial n_{k'}^0}{\partial \epsilon_{k'}^0}\nu_{k'}=\int_{k'}f_{kk'}\delta n_{k'}=0
\end{align}

\noindent The latter occurs when $\nu_k=0$, and is thus the non-physical solution. For the former solution, we find that 

\begin{align}
\nu_k&=\frac{\cos\theta}{s-\cos\theta}\left(
\frac{\int_{k'}f_{kk'}\frac{\partial n_{k'}^0}{\partial \epsilon_{k'}}\nu_{k'}}{
1+\frac{1}{4\alpha v_F}\left(\frac{1}{\cos\theta-s}\right)\int_{k'}f_{kk'}\frac{\partial n_{k'}^0}{\partial \epsilon_{k'}}a_{kk'}^s
}
\right)\notag\\
&=\frac{\cos\theta}{s-\cos\theta}\left(
\frac{\int_{k'}f_{kk'}\frac{\partial n_{k'}^0}{\partial \epsilon_{k'}}\nu_{k'}}{
1+ \frac{s}{\cos\theta-s}
}
\right)\notag\\
&=-\int_{k'}f_{kk'}\frac{\partial n_{k'}^0}{\partial \epsilon_{k'}}\nu_{k'}
\end{align}
\added{This is essentially a linear, homogenous Volterra equation for the function $\nu_k$. We can map this to a linear homogenous second order differential equation, where one of the solutions is an exponential as a function of $k$. The other solution is for $\nu_k=0$, which is what we see in a conventional Landau-Fermi liquid when $s=0$. However, because $\nu_k$ experiences a singularity at this point, $\nu_k$ cannot be zero (as stated before), and thus the Majorana-Fermi surface distortion must increase exponentially for the above values. In a Landau-Fermi liquid, such divergences appear when $s$ is in the particle-hole channel, however we avoid such a change in topology (known formally as a Lifshitz transition \cite{Lifshitz}) by assuming an angle-dependent interaction and subsequently performing a simple contour integration. This leads to the prediction of Landau damping in the Landau-Fermi liquid. In the Landau-Majorana liquid, however, the robust stability of the Majorana-Fermi surface "lifts" the Lifshitz transition above the particle-hole channel.}

\section{V. Zero Sound in a Landau-Majorana liquid} From the above analysis, it appears that, although the Landau-Majorana liquid is highly stable, an increase in effective screening brought about by stronger repulsive interactions leads to a highly unstable Lifshitz transition, after which the reality condition is completely suppressed and Fermi liquid behavior is restored. To explore the system further, we derive explicitly the interaction-dependence of the Landau-Majorana liquid zero sound.

The zero-sound in the screened Landau-Majorana system is described by an equation similar to that describing the Landau-Fermi liquid zero sound\cite{Landau2,Mermin}, except now we have an additional term dependent on the scattering amplitude $A_0^s=a_0^s N(0)$\footnote{We take $F_0^s=N(0)f_0^s$ and $A_0^s=N(0)a_0^s$ here and throughout this article, but we can replace the results with $F_0^a$ or $A_0^a$ without loss of generality.}:
\begin{align}
1+\frac{1}{2}\log \left(\frac{s-1}{s+1}\right)\left\{s+\frac{A_0^s}{\widetilde{v}}\right\}=-\frac{1}{F_0^s}\label{eq18}
\end{align}
\\
\noindent \added{where the unitless parameter $\widetilde{v}$ is given by}
\begin{align}
\widetilde{v}=
& 4 \alpha \left(\frac{k_F}{\pi}\right)^{d-1} \Omega
\end{align}
\added{the scattering amplitude $A_0^s$ is}
\begin{align}
\added{A_0^s=\frac{F_0^s}{1+F_0^s} }
\end{align}
\added{and where $\Omega$ is the real space volume in $d\ge 2$ spatial dimensions. Note that the term proportional to this factor is the direct result of the hole suppression inherent to the Majorana polaron system and utilized in the derivation of the Landau-Majorana kinetic equation Eqn. \eqref{eq8} (See Appendix B for derivation). We first solve the above equation for  $A_0^s/\widetilde{v}>>s$ (i.e., we are either near the $F_0^s\sim -1$ or $\alpha\rightarrow 0$ limits). In such a scenario, we find that}
\begin{align}
s\approx \begin{cases}
\tanh\left(\frac{\widetilde{v}}{(A_0^s)^2}\right),\quad \left|\frac{A_0^s}{\widetilde{v}}\right|>>|s|,\quad s<0\\\ \\
\coth\left(\frac{\widetilde{v}}{(A_0^s)^2}\right),\quad \left|\frac{A_0^s}{\widetilde{v}}\right|>>|s|,\quad s>0 \label{eq32}
\end{cases}
\end{align}

\begin{figure}[t]
 \begin{center}
\includegraphics[width=.8\columnwidth]{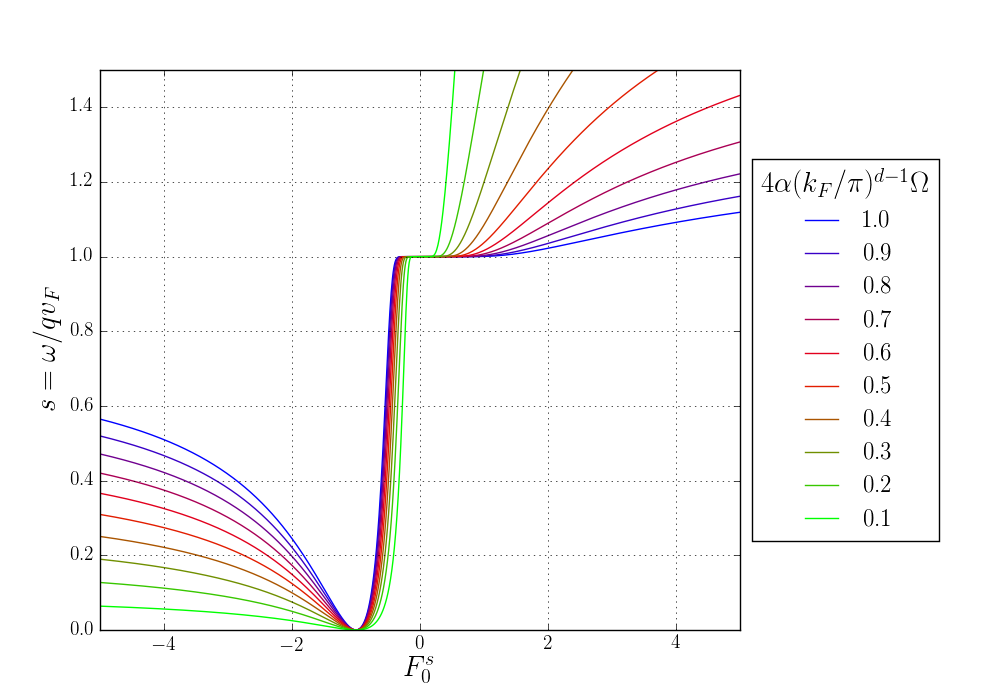}
\captionof{figure}{
\added{The dimensionless zero sound dispersion $s=\omega/qv_F$ vs. Landau parameter $F_0^s$ for several values of $\widetilde{v}$, with the assumption that we are in the limit of $|A_0^s/\widetilde{v}|<<|s|$. For small $\widetilde{v}$, the zero sound dispersion drops to zero before the usual $F_0^s=-1$ instability, while it is marginally revived for $F_0^s<-1$ as it begins to saturate to a value of $s\sim \tanh(\widetilde{v})$. For larger $\widetilde{v}$, the dispersion saturates to a value that slowly approaches $1$, leading to a breakdown of the assumption considered in the derivation of Eqn. \eqref{eq32} and the resulting behavior in the above figure becoming unphysical away from the point $F_0^s=-1$. Also note that there is a breakdown of our assumption in the derivation of Eqn. \eqref{eq32} in very close proximity to $F_0^s=0$, leading to the unphysical flat portion of the dispersion in the above figure.}}\label{fig:Fig2}
\end{center}
\end{figure}

\added{A plot of this dispersion is given in Fig \ref{fig:Fig2}. Although this is only valid for $\alpha\rightarrow 0$ and small $s$ (i.e., small $F_0^s$), the behavior observed is significantly different from the regular Landau-Fermi liquid result. In the traditional Landau-Fermi liquid, an approximate analytical solution for $s$ is confined to $|s|\sim 1$. In our case, we have a closed form for all $s$ as long as we take the Majorana parameter to be very, very small. More importantly, the Landau-Fermi liquid for Landau parameters $-1<F_0^s<0$ is characterized by a finite imaginary component of the dispersion. For $F_0^s<-1$, the dispersion becomes purely imaginary and thus critically damped. For the Landau-Majorana liquid in the "Majoranic limit", however, we see that $F_0^s<0$ is characterized by a real dispersion. For $\widetilde{v}<<0$, the dispersion approaches zero very fast (before we reach the Pomeranchuk instability proper for a traditional Fermi liquid \cite{Pomeranchuk}), and then increases to some finite value. As $\widetilde{v}$ grows, then $s$ in the regime $F_0^s<-1$ grows as well, however this behavior violates our original assumtion in the derivation of Eqn. \eqref{eq32}. Therefore, we find that the zero sound dispersion of the Landau-Majorana liquid with small $\widetilde{v}$ has a value of $s\sim 1$ for $F_0^s\sim 0$, a value of $s\sim 0$ for $F_0^s\sim -1$, and a small but finite value for $F_0^s<-1$. This behavior is only seen for $\widetilde{v}<<0$; beyond this case, we need to take a different approximation.}

\begin{figure*}[htbp]
\hspace*{-10mm} 
\begin{subfigure}{.55\columnwidth}
\includegraphics[width=1\columnwidth]{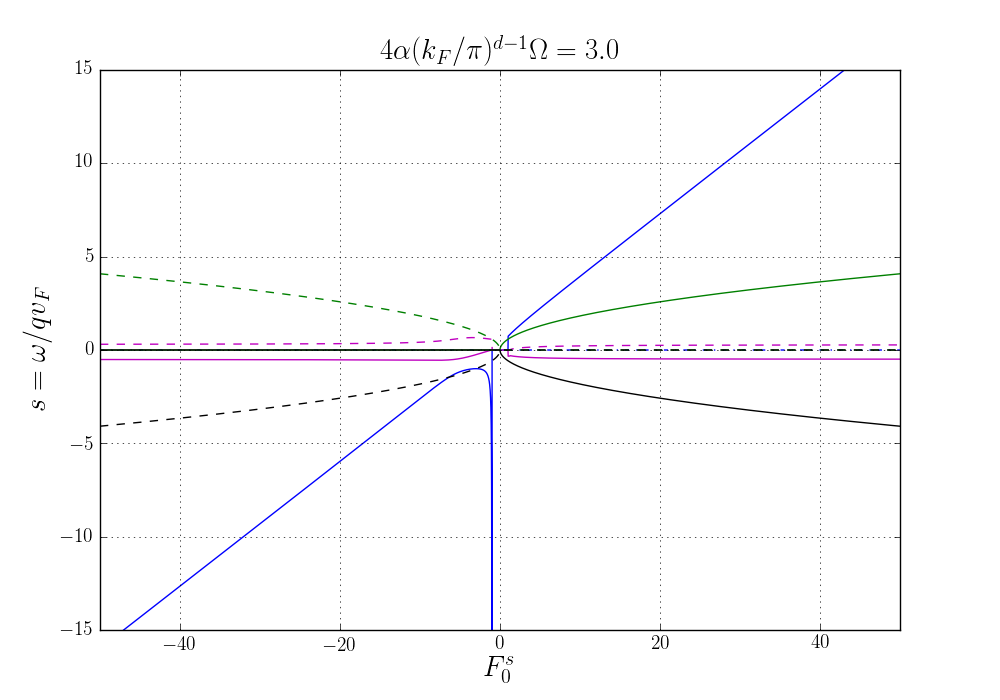}
  \caption{$\widetilde{v}\sim 3.0$}\label{fig:2a} 
\end{subfigure}
\begin{subfigure}{.55\columnwidth}
\includegraphics[width=1\columnwidth]{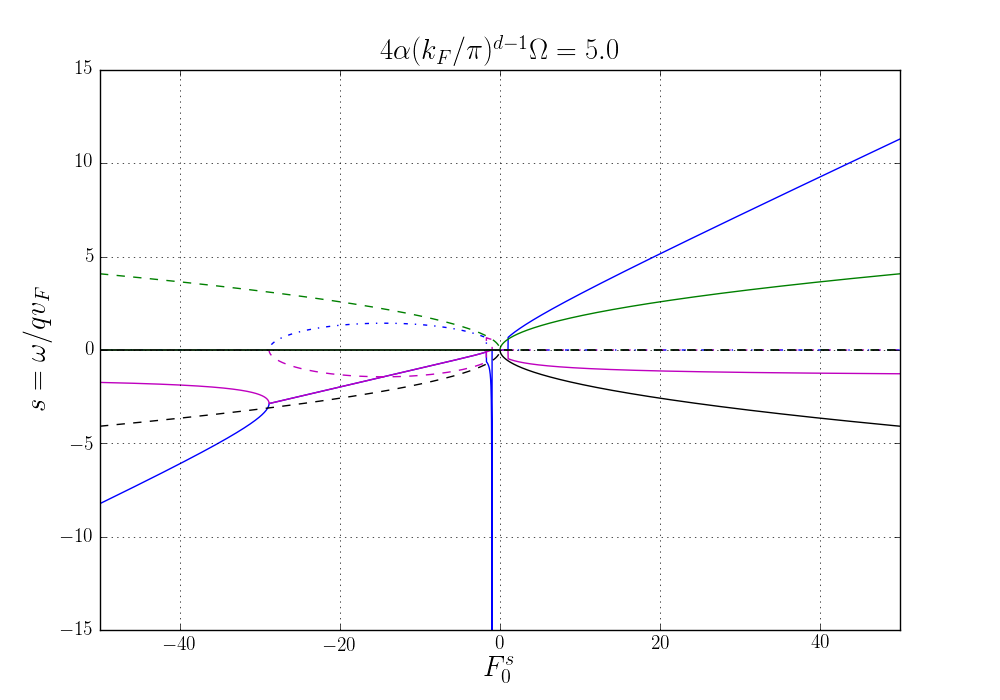} 
  \caption{$\widetilde{v}\sim 5.0$}\label{fig:2b} 
\end{subfigure}
\hspace*{-10mm} 
\begin{subfigure}{.55\columnwidth}
 \includegraphics[width=1\columnwidth]{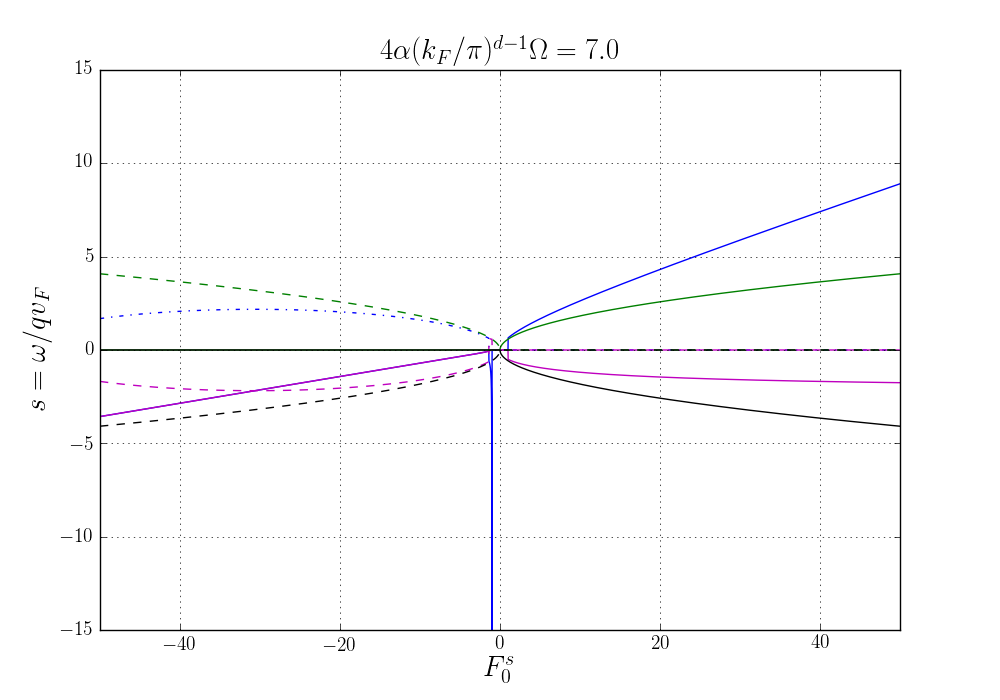} 
   \caption{$\widetilde{v}\sim 7.0$}\label{fig:2c} 
\end{subfigure}
\begin{subfigure}{.55\columnwidth}
\includegraphics[width=1\columnwidth]{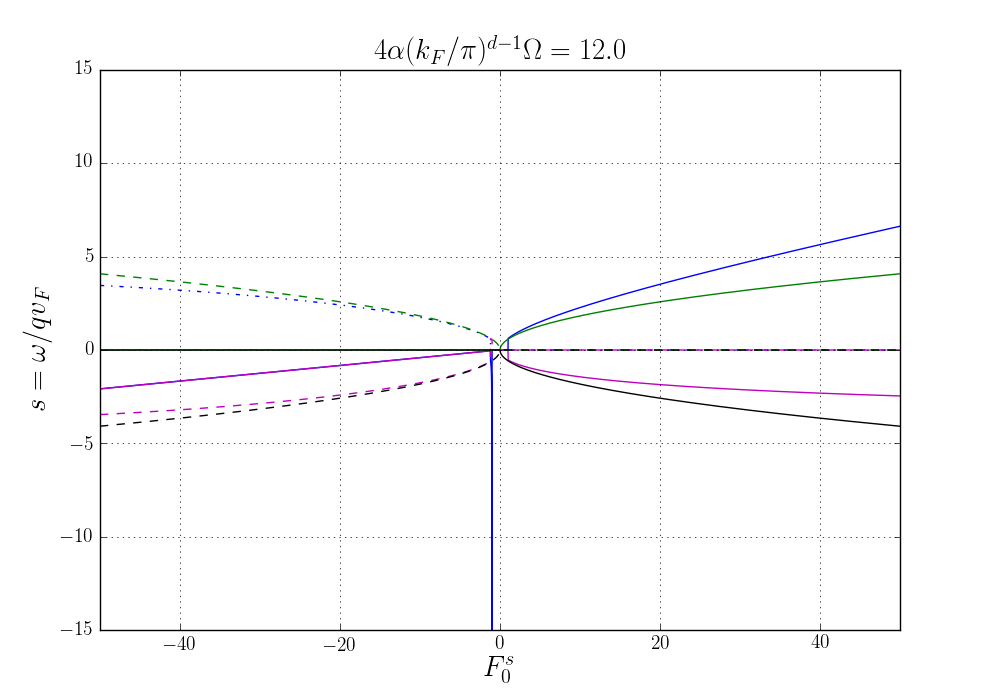} 
   \caption{$\widetilde{v}\sim 12.0$}\label{fig:2d} 
\end{subfigure}
\caption{{\color{black} 
Behavior of the zero sound dimensionless parameter $s$ vs. interaction strength $F_0^s$ \added{for several values of $\widetilde{v}$. The solid (dotted) green lines are the real (imaginary) parts of the positive branch of the Landau-Fermi liquid zero sound dispersion, while the solid (dotted) black lines are the real (imaginary) parts of the negative branch of the Landau-Fermi liquid result. In contrast, the solid (dotted) blue line is the real (imaginary) component of the first branch of the Landau-Majorana zero sound dispersion, while the pink line is the other Landau-Majorana branch. The singularity seen in the former for small $F_0^s$ is an artifact of Eqn. \eqref{main}'s breakdown for weak interaction, and is unphysical. Note that the negative branch of the Landau-Fermi liquid dispersion is unphysical for $F_0^s>0$ because it's completely real and negative, while the positive branch of the Fermi liquid dispersion is unphysical for $F_0^s<0$ due to causality arguments. In a similar fashion, the pink dispersion of the Landau-Majorana liquid result is unphysical for $F_0^s>0$, while the blue dispersion is unphysical for $F_0^s<0$ for the same reasons. For $F_0^s$ below the "bifurcation point", only two real negative dispersions may exist, leading to an instability in the fluid state.}} }
\label{fig:Fig3}
\end{figure*}

\added{To consider the case of $\widetilde{v}\sim 1$ (which would correspond to a larger $\alpha$ or larger $\Omega$), we will need to consider a different limit of Eqn. \eqref{eq18}.} 
%
%
%
%
%
 Assuming $|F_0^s|>>1$ leads to $|s|>>1$ (as in the Landau-Fermi liquid), Eqn. \eqref{eq18} simplifies to
\\
 \begin{align}
\frac{1}{F_0^s}&=\frac{1}{3s^2}+\left(\frac{F_0^s}{3(1+F_0^s)\widetilde{v}}\right)\frac{1}{s^3}
+\left(\frac{F_0^s}{(1+F_0^s)\widetilde{v}}\right)\frac{1}{s}\label{main}
\end{align}
\added{In the limit of $\widetilde{v}\rightarrow \infty$, we recover the well-known zero sound dispersion in a regular Landau-Fermi liquid:}
\begin{align}
\added{s=\sqrt{\frac{F_0^s}{3}}}
\end{align}
\\
\noindent \added{In the case of the Landau-Majorana system with finite $\widetilde{v}$, the full solutions to Eqn. \eqref{main}} are rather lengthy, and are reproduced in full in Appendix B. There are three equations for the value of velocity $s$, each with differing dependence on the interaction strength $F_0^s$. This dependence is shown in Fig. \ref{fig:Fig3}.

%

For large positive values of $F_0^s$, there are two possible dispersions corresponding to an upper and lower branch of the excitation spectrum. The upper branch behaves similar to a regular Landau-Fermi liquid, growing larger with larger values of $F_0^s$ and living in the $s>0$ regime. The lower branch corresponds to $s<0$, and converges to a finite value as $F_0^s\rightarrow \infty$. \added{As the latter branch is purely real and only exists for $\omega<0$ (with $qv_F$ taken to be positive definite), we interpret this as an unphysical branch.}
%
%
%
%
%
%
%
\added{Below $F_0^s=0$, we observe the coexistence of a real and imaginary portion to the dispersion. This regime corresponds to the Landau damped regime,
where the imaginary part of the renormalized zero sound velocity in a Landau-Majorana liquid dies at a particular \added{large} value of attractive $F_0^s$, after which the real part of $s$ bifurcates into $s_1$ and $s_2$. As both of the modes below the bifurcation point for negative $F_0^s$ are real and defined by $\omega<0$, we interpret these unphysical, purely real dispersions beyond the second bifurcation to be a sign of the breakdown of the Landau-Fermi liquid like ground state.}

\added{It is important to compare this work with the predictions of traditional Landau-Fermi liquid theory. The well-known Landau kinetic equation corresponds to Eqn. \eqref{eq18} with $\alpha\rightarrow \infty$. Assuming that $s=\omega/qv_F$ is complex (i.e., $s=s'+is''$, $s',\,s''\not=0$), then $s'$ is allowed to be positive or negative, but is ultimately constrained to be below unity as we are bounded to the condition that $|s|<1$. A value of $|s|>1$ for complex zero sound dispersion would correspond to $F_0^s<-1$, which violates the Pomeranchuk instability condition. This makes sense, as a
"downturn" in the sound velocity is generally considered the telltale sign of a localization transition, where the resultant soft mode causes a dip in the dispersion as $q$ becomes on the order of the reciprocal lattice vector \cite{Roton}.
%
%
%
 In a Landau-Fermi liquid, however, such a strong "negative" sound dispersion is inaccessible.}

\added{
In the Landau-Majorana liquid, recall that the Landau-damped regime is extended beyond the small $|s|$ (i.e., small $F_0^s$) region, as evident from the finite real component in Fig. \ref{fig:Fig3} and Eqn. \eqref{screened}, where singularities of the zero sound eigenvector occur when 
$\cos\theta=s-\frac{1}{\gamma}$, where $\gamma\sim \frac{\alpha}{F_0^s A_0^s}$. As stated before, the regime $F_0^s>1$ will ensure $\gamma>0$, and thus such singularities will be "lifted" above the particle-hole channel. However, for $F_0^s<-1$, we have $\gamma<0$, implying that such damped modes must correspond to a large, negative value of $s'$. Interestingly, as $F_0^s$ continues to decrease, our analysis shows that $s'$ will continue to become more negative while $s''$ approaches zero. Therefore, for sufficiently negative interactions and a small but finite value of $\alpha$, we see that $\cos\theta\sim s'-\frac{1}{\gamma}<0$, or in other words there exists a nearly-dispersionless mode in a system characterized by $\epsilon_{q+p}-\epsilon_p<0$, where $\epsilon_p$ is the dispersion of the single-particle spectrum. This is highly different from the usual Landau-Fermi liquid case, where a weakly negative $s'$ coexists with a strong damping term. This leads to two possibilities: either the Landau-Majorana liquid is unstable directly before it hits the Pomeranchuk instability proper, or we have a stable Fermi-liquid like state  if the spectrum of quasiparticle excitations is negative. }

\added{
The study of collective zero sound modes outlined in this section leads to an experimental hallmark of the Landau-Majorana liquid in the presence of weakly attractive interactions: a sound dispersion at the Fermi surface with a strong negative real component and, near the unstable bifurcation point, a highly stable Fermi-liquid phase if we have a "back-bending" dispersion for single-particle excitations. Although usually present in superconducting systems with a finite energy gap, a back-bending dispersion has been seen experimentally in the pseudogap phase of the underdoped cuprates above $T_c$ \cite{Tu,Kanigel,Hashimoto} and strongly interacting Fermi gases in the normal state \cite{Gaebler, Feld}. This has led to the proposal that the emergent "back-bending" dispersion of single-particle excitations might be a universal signature of independently excited particles in the presence of residual condensation \cite{Kanigel,Qijin,Chen_Levin,Chen_2009}. Moreover, such back-bending has been proposed to be intimately linked with a non-zero spectral weight below the Fermi surface for $k>>k_F$ and, consequently, a "sharp" drop in the momentum distribution \cite{Schneider}, echoing what we have already seen in the Majorana polaron statistics.
Our prediction of a back-bending dispersion in the presence of a stable Majorana-Fermi surface (as well as the tendency towards superconductivity discussed in the next section) suggests that the Landau-Majorana liquid theory could extend the quasiparticle paradigm to such unconventional phases in the absence of Fermi arcs \cite{Proust}, complementing existing studies of "Luttinger's theorem-violating Fermi liquids" \cite{Mei, Sachdev, Chatterjee} by extending the Landau paradigm to quasiparticle-like objects already observed in certain underdoped cuprates \cite{Vishik,Grissonnanche}. Of course, we acknowledge that pseudogap phenomenon is rich and complex, and that future experiments must be done on such materials to see to what extent they can be modeled as a Landau-Majorana liquid. 
}
%
%
%
%

\section{VI. Pomeranchuk instability in the Landau-Majorana liquid and experimental predictions}

\added{\hspace{9mm}As a finite real component of the dispersion remains for a reasonably large regime of negative $F_0^s$, the main results in the previous section tell us that the Landau-Fermi liquid is stable against Pomeranchuk instabilities for a large class of attractive interactions. In a traditional Landau-Fermi liquid, a Pomeranchuk instability is the spontaneous development of either long-range spin or charge order at some critical Landau parameter  $F_\ell^s=-(2\ell+1)$, at which point a phase separation occurs and the Fermi liquid becomes more akin to a liquid crystal phase \cite{Fradkin1, Fradkin2}. This instability condition can be reproduced microscopically, where the quasiparticle-quasihole scattering amplitude can be directly calculated from the Bethe-Salpeter equation:}
\begin{align}
\Gamma_\ell^s=\frac{F_\ell^s}{1+F_\ell^s/(2\ell+1)}
\end{align}
\added{The prefactor is found in the above by summing over particle-hole bubbles in a geometric series. One can then conclude that the Pomeranchuk instability condition is the direct result of quasiparticle-quasihole scattering, which results in 
in an exponential accumulation of density fluctuations at the Fermi surface. Nevertheless, there have several attempts to circumvent such instabilities by either considering a local approximation to the self-energy \cite{Engelbrecht} or by considering a static Bose condensate of scalar modes \cite{Kolomeitsev}. In both cases the particle-hole vertex is significantly altered, yielding a novel renormalization of the Landau parameters and rendering the liquid free from the Pomeranchuk condition. In the system we consider in this paper, we find a retarded quasihole excitation in the Landau-Majorana liquid as the direct consequence of invoking the "sharp" Fermi surface of the Majorana polaron statistics. The result of introducing the $\alpha$-parameter characterizing the Majorana-Fermi surface results in a new "effective" interaction $\widetilde{F}_\ell^s$ (shown in Appendix A), much as in \cite{Engelbrecht, Kolomeitsev}.
Therefore, a reduced hole contribution from the underlying Majorana polaron statistics will then result in a reduced number of quasiparticle-quasihole scattering events in the fermionic degrees of freedom, making the Landau-Majorana liquid more stable than the traditional Landau-Fermi liquid. This agrees with recent theoretical work of neutral spinon Fermi surfaces in Mott insulators, where local pairing instabilities are suppressed in the presence of a sharp Fermi surface with no conventional Landau-Fermi liquid \cite{Max,Shen}. The main difference between this work and the theory outlined in this article, however, is that the Landau quasiparticle concept may still be applied in our system as long as we adiabatically evolve from the non-interacting Majorana polaron gas.}

\added{
As the regime $F_0^s$ below the first bifurcation in Fig. \ref{fig:Fig3} is unphysical, we are tempted to take such a point as the true onset of a Pomeranchuk instability in the Landau-Majorana liquid. We can check this by finding the exact Landau parameter which corresponds to this bifurcation, and comparing it with the compressibility of the Landau-Majorana liquid. Two points of bifurcation are found for $F_0^s<0$ in Appendix B, which correspond to the densities $n_c^{(1)}$ and $n_c^{(2)}$:
}
\begin{equation}
n_c^{(1)}\approx \frac{\pi}{d}\left(
\frac{\sqrt{-3F_0^s}}{8\alpha \Omega}\right)^{\frac{d}{d-1}}\label{eq22a},\quad
\quad 
n_c^{(2)}\approx \frac{\pi}{d}\left(\frac{\sqrt{3}}{2\alpha \Omega}\right)^{\frac{d}{d-1}}
\end{equation}
\added{The first of the above corresponds to the bifurcation which we predict to signal the onset of a Pomeranchuk instability condition. This can be easily seen by looking at the form in the above: when $F_0^s\sim -n^{2\left(\frac{d-1}{d}\right)}\alpha^2 \Omega^2$, then all Landau parameters below this value will correspond to two real zero sound dispersions that are completely negative, and hence unphysical. The second of the above, however, is somewhat surprising, as it occurs at low densities and is independent of interaction. As the above discussion only holds for systems at high densities (i.e., sizable screening), and this bifurcation only occurs in the dilute limit, we interpret this value as the lower limit of density at which our screening ansatz makes sense, as collective excitations at densities $n<n_c^{(1)}$ would be Landau damped regardless of interaction strength and therefore be unphysical.}

{Rearranging the expression for $n_c^{(1)}$, we find that the value of the Landau parameter $F_0^s$ at the point of bifurcation is given by}
\begin{align}
F_0^s=-\frac{64}{3} \alpha^2 \Omega^2 \left(\frac{k_F}{\pi}\right)^{2(d-1)}\label{35}  
\end{align}
\added{
Therefore, for any interaction $F_0^s<0$ with magnitude larger than the above, the system appears to become unstable. As stated before, we can check this by looking at the compressibility of the Landau-Majorana liquid. This is derived in Appendix A, and reproduced below:
}

\begin{align}
\widetilde{\kappa}&=\frac{1}{n^2}\frac{\partial n}{\partial \widetilde{\mu}}\bigg|_T\approx\frac{4}{n^2}\left(\frac{N^*(0)}{1+\frac{k_F}{4\alpha}+F_0^s}\right)\label{eq23}
\end{align}
%
The compressibility is positive as long as $F_0^s>-\left(1+\frac{k_F}{4\alpha}\right)$. \added{For $F_0^s<-\left(1+\frac{k_F}{4\alpha}\right)$, the compressibility is negative, signaling an instability. We see similar behavior from Eqn. \eqref{35}, derived from the bifurcating zero sound dispersion; i.e., $\alpha\rightarrow 0$ results in an increased stability for large attractive interaction. Interestingly, from the above comparison, we are able to derive a closed from for the ratio $\alpha/k_F$ assuming $\alpha/k_F<<1$}:
\begin{align}
\frac{\alpha}{k_F}\sim \left(\frac{3\pi^{2(d-1)}}{256}\right)^{1/3} \left(\Omega k_F^d\right)^{-2/3}\label{38}
\end{align}
%
%
%
\added{Because we expect both the prediction from the zero sound dispersion and the compressibility to agree, Eqn. \eqref{38} tells us $\alpha$ should disappear in the thermodynamic limit, agreeing with the previous discussions. Moreover, the above example allows us to make theoretical predictions for real materials. As stated previously, a likely candidate material for the Landau-Majorana liquid is the Kondo insulator SmB$_6$, due to the proposal of a Fermi surface of neutral fermionic excitations with a small, robust Fermi surface in the bulk \cite{Baskaran,Pixley}. Using quantum oscillation data taken in this material near the 3D limit \cite{Sebastian}, we find a value of $\alpha/k_F\sim 10^{-13}$, where $k_F$ is the Fermi wavevector for the bulk Fermi surface proposed in this material. From similar quantum oscillation data taken on underdoped YBa$_2$Cu$_3$O$_{6.5}$ in a high magnetic field \cite{Proust} (i.e., a quasi-2D high-$T_c$ superconductor in the pseudogap phase with disappearing Fermi arcs), we find values of $\alpha/k_F\sim 10^{-6}$. The former is an example of a possible 3D Majorana parameter, while the latter is a possible 2D Majorana parameter. If the excitations in the Kondo insulator and underdoped cuprate can both be modeled as a Landau-Majorana liquid, a direct experimental probe of the momentum distribution function of such fermionic quasiparticles should then yield a "sharp" Fermi surface parameterized by these values of $\alpha/k_F$. Unfortunately, such values of $\alpha/k_F$ in these experiments are not ideal for the verification of our zero sound predictions: for the experiments done on SmB$_6$, we find that $\widetilde{v}\sim 10^6$, while for the YBa$_2$Cu$_3$O$_{6.5}$ data we obtain $\widetilde{v}\sim 10^2$. Both of these values would correspond to the Landau-Fermi liquid like dispersions, with a negligible real component for the zero sound dispersion below $F_0^s=-1$. To see a more robust zero sound mode in the Landau-Majorana liquid, the proposed system would have to be a few orders of magnitude smaller than current experiments, preferably in the 2D limit. Nevertheless, it is important to note that the dispersion described in Eqn. \eqref{eq32} is still valid for large values of $\widetilde{v}$ if we are close in proximity to $F_0^s=-1$. A finite real dispersion in close proximity but below the conventional Pomeranchuk instability point in these materials would be a tell-tale sign of Landau-Majorana behavior.
}

\added{The above discussion tells us something very important: namely, that coherent Majorana-like excitations in the finite low-temperature limit will be more stable to pairing instabilities than their complex counterparts, irrespective of the underlying material. A modified Pomeranchuk instability in systems with Majorana excitations is most likely to be experimentally detected in the compressibility or effective mass of such quasiparticles. As stated before, the compressibility diverges at the point of phase separation in the $\ell=0$ channel, while the effective mass disappears at the instability in the $\ell=1$ channel. In a Landau-Majorana liquid, we expect from the forms of the compressibility and effective mass that the $\ell=1$ Pomeranchuk instability to be unchanged from the Landau-Fermi result(see Appendix A), yet the $\ell=0$ channel to be much more robust. 
To directly test our prediction, an ideal experiment would explore the robustness of a Majorana-Fermi surface to distortion in the screened, collisionless regime. Of particular interest are the spin-orbit entangled lithium iridates, which have been predicted to harbor stable Majorana-Fermi surfaces \cite{Hermanns}. Recent experiments on the silver-lithium iridate Ag$_3$LiIr$_2$O$_6$ \cite{Tafti} have shown that a metallic-like phase of itinerant Majorana fermions exists in a finite-temperature regime of the Kitaev magnet with suppressed antiferromagnetic order, in agreement with unbiased Quantum Monte Carlo simulations on the honeycomb lattice \cite{Nasu}. To check its robustness against Fermi surface deformation, one might explore the range of superconducting instabilities in this material. In similar compounds like Li$_2$IrO$_3$, it has been proposed that the underlying $Z_2$ topological order that results in Majorana excitations (and of which our work is a finite-temperature extension) breaks down in the presence of hole doping, and a p-wave superconducting instability emerges in the absence of Heisenberg coupling \cite{Ashvin, Rosenow}. 
As a finite superconducting gap suppresses Fermi surface deformations \cite{Halboth, Yamase, Barci}, the presence of a nearly universal superconducting instability for strong attractive interactions would signal the absence of a Pomeranchuk instability, and would therefore be in agreement with our theory. To put it more succinctly, the only pairing that should occur in a Landau-Majorana liquid should be through the particle-particle channel. The exact nature of such a pairing mechanism in the Landau-Majorana liquid, however, is beyond the scope of the current work, as it would move well beyond the Landau quasiparticle paradigm.}
\section{VIII. Conclusions} \added{In several materials, there is strong evidence of a neutral fermionic fluid-like state that defies a traditional description in terms of Landau-Fermi liquid theory.} In this paper, we have \added{proposed an extension of the Landau-Fermi liquid formalism} to a general system of interacting fermions obeying a self-conjugacy relation. This ultimately amounts to a Landau quasiparticle extension of the previously established Majorana polaron model \cite{Heath}. At zero temperature, the equilibrium system mirrors the non-interacting system of complex fermions. However, as we perturb the system, the suppression of the quasihole excitations in the Majorana-Fermi sea \added{(a direct consequence of the Majorana polaron statistics)} manifests itself as an amplification of the Landau-Majorana quasiparticle effective mass. By introducing the \added{Majorana parameter} $\alpha$, \added{we can "tune the sharpness" of the Fermi surface, allowing us to define a smooth transition from Landau-Majorana to Landau-Fermi like behavior. In between the two phases, we find that the system is characterized by a Lifshitz transition.} The interaction dependence of the dimensionless zero sound parameter $s=\omega/qv_F$ shows \added{a robust stability of the underlying Landau-Majorana liquid state unseen in the Landau-Fermi liquid case, agreeing with a calculation of the Pomeranchuk instability.} 

We believe such a study provides a simple proposed model of highly complex materials with unconventional quasiparticle excitations, such as Kitaev spin liquids \cite{Yiping, Hermanns,Takikawa}, the Kondo insulator SmB$_6$\cite{Tan,Sebastian}, \added{and possibly certain pseudogap materials in the limit of large magnetic field \cite{Grissonnanche,Proust}}, where recent theoretical proposals and experimental evidence points to Landau-Fermi liquid-like behavior without adiabatic continuity to the non-interacting Fermi gas. The Landau-Majorana liquid proposed here is a rigorous attempt to understand these materials with the well-established formalism of Landau-Fermi liquid theory \added{by adiabatically connecting the interacting phase with a novel statistical theory with self-conjugate fermionic degrees of freedom. Ultimately, our work} provides the first known attempt to derive universal experimental signatures of many-body Majorana physics via a description of the $\ell=0$, zero-temperature acoustic modes \added{and Fermi surface instabilities}.
\\\\
{\it Acknowledgements.} The authors thank Faranak Bahrami, Kenneth Burch, Caitlin Duffy, Hsu Liu, Natalia Perkins, Fazel Tafti, and Yiping Wang for exceptionally useful discussions on the experimental detection of Majorana fermions in several different materials. 
 This work was partially supported by the John H. Rourke Endowment Fund at Boston College. 


\section{Appendix A: Derivation of the Landau-Majorana compressibility and effective mass}
We may now derive the Landau-Majorana compressibility. In general, the compressibility is given by
\begin{align}
\kappa=\frac{1}{n^2}\frac{\partial n}{\partial \widetilde{\mu}}\bigg|_T
\end{align}
We now write down the expression for the partial functional derivative of the Landau-Majorana liquid chemical potential:

\begin{align}
\frac{\partial \widetilde{\mu}}{\partial n}=\frac{\partial \widetilde{\epsilon}_k}{\partial k}\frac{\partial k}{\partial n}\bigg|_{k_F}+\frac{1}{V}\sum_{k'\sigma'}\widetilde{f}_{\sigma \sigma'}(k,\,k')\bigg|_{k_F}\frac{\partial n_{k'}}{\partial k}\frac{\partial k}{\partial n}\bigg|_{k_F}
\end{align}
Where
\begin{align}
\frac{\partial \widetilde{\epsilon}_k}{\partial k}\bigg|_{k_F}&=\frac{\partial}{\partial k}(\epsilon_k \Theta(k-k_F))\bigg|_{k=k_F}\notag\\
&=\epsilon_k\delta(k-k_F)\bigg|_{k_F}+\frac{\partial \epsilon_k}{\partial k}\Theta(k-k_F)\bigg|_{k_F}\notag\\
&\approx \epsilon_k \frac{\mathfrak{F}(k_F,\,\alpha)
(1-\mathfrak{F}(k_F,\,\alpha)
)}{\alpha}+\frac{\partial \epsilon_k}{\partial k}\mathfrak{F}(k_F,\,\alpha)
\end{align}

\noindent \added{In the above, note that it is the approximation $\widetilde{\epsilon}_k\equiv \epsilon_k \Theta(k-k_F)$ which contains the physics of suppressed quasi-hole states introduced in Eqn. \eqref{eq5}.}
 Note that $\mathfrak{F}(k,\,\alpha)=1$ if we take $k-k_F>>\alpha$ and $\mathfrak{F}(k,\,\alpha)=1/2$ for $k-k_F<<\alpha$. Hence,
\begin{align}
\frac{\partial \widetilde{\epsilon}_k}{\partial k}\approx\begin{cases} \frac{\partial \epsilon_k}{\partial k}\bigg|_{k=k_F},\quad &k-k_F>>\alpha\notag\\
\frac{\epsilon_{k_F}}{4\alpha}+\frac{1}{2}\frac{\partial\epsilon_k}{\partial k}\bigg|_{k=k_F},\quad &k-k_F<<\alpha
\end{cases}
\end{align} 
If we take the latter case, then
\begin{align}
\frac{\partial \widetilde{\epsilon}_k}{\partial k}\bigg|_{k_F}&\approx\frac{\epsilon_{k_F}}{4\alpha}+\frac{v_F}{2}\notag\\
&=\frac{k_F^2}{8m^* \alpha}+\frac{k_F}{2m^*}\notag\\
&=\frac{k_F}{2m^*}\left(1+\frac{k_F}{4\alpha}\right)\notag\\
&=\frac{v_F^*}{2}\left(1+\frac{k_F}{4\alpha}\right)\notag\\
&\equiv v_M^*
\end{align}
Where we define for convenience $v_M^*$ as the Landau-Majorana velocity. 

It is interesting to note that, from the above form of the Landau-Majorana velocity, we could effectively define a Majorana density of states, and from this obtain the Landau-Majorana specific heat. Much as in the non-interacting Majorana polaron gas\cite{Heath}, the specific heat $C_v$ for the Majorana system has the same temperature-dependence as the Fermi-Dirac case. Because the temperature-dependence of $C_v$ doesn't change as we go from a Landau-Fermi liquid to the Landau-Majorana liquid, Luttinger's theorem is still satisfied in the latter, and thus the Landau-Fermi liquid picture is applicable to our system.\cite{Haldane}.

The other quantities in the expression for $\frac{\partial \widetilde{\mu}}{\partial n}$ are identical to the Landau-Fermi liquid. We then find that

\begin{align}
\frac{\partial \widetilde{\mu}}{\partial n}&=\frac{\partial \widetilde{\epsilon}_k}{\partial k}\frac{\partial k}{\partial n}\bigg|_{k_F}+\frac{1}{V}\sum_{k'\sigma'}\widetilde{f}_{\sigma \sigma'}(k,\,k')\bigg|_{k_F}\frac{\partial n_{k'}}{\partial k}\frac{\partial k}{\partial n}\bigg|_{k_F}\notag\\
&\approx\frac{\pi^2}{k_F^2}\left\{
v_M^*+\frac{k_F^2}{2\pi^2}f_0^s
\right\}\notag\\
&=\frac{\pi^2}{k_F^2}\left\{v_M^*+N^*(0)v_F^*f_0^s\right\}\notag\\
&\approx\frac{1+\frac{k_F}{4\alpha}+F_0^s}{4N^*(0)}
\end{align}
where $N^*(0)=\frac{m^* k_F}{2\pi^2}$. This leads directly to Eq. \eqref{eq23}.

We now move onto the modified Landau parameter for $\ell>0$. Interestingly, we can write the Majorana compressibility derived above in the form
\begin{align}
\kappa&=\frac{2}{n^2}\left(\frac{\widetilde{N}^*(0)}{1+\widetilde{F}_0^s}\right)
\end{align}
where we have defined
\begin{align}
\widetilde{N}^*(0)=2N^*(0)
\end{align}
and
\begin{align}
\widetilde{F}_0^s=\frac{k_F}{4\alpha}+F_0^s
\end{align}
Looking at the interaction term, we see that
\begin{align}
\widetilde{f}_0^s&=\frac{k_F}{4\alpha N^*(0)}+f_0^s\notag\\
&=\frac{k_F}{4\alpha N^*(0)}+\frac{1}{2}\int_{-1}^1d(\cos\theta)f_{kk'}^s\notag\\
&=\frac{1}{2}\int_{-1}^1 d(\cos\theta)\widetilde{f}_{kk'}^s
\end{align}
where we have now defined 
\begin{align}
\widetilde{f}_{kk'}^s =\frac{k_F}{4\alpha N^*(0)}+f_{kk'}^s
\end{align}
With this form, we can then see that
\begin{align}
\widetilde{f}_1^s&=\frac{3}{2}\int_{-1}^1 d(\cos\theta)\cos\theta \widetilde{f}_{kk'}^s \notag\\
&=\frac{3}{2}\int_{-1}^1 d(\cos\theta)\cos\theta \frac{k_F}{4\alpha N^*(0)}+\frac{3}{2}\int_{-1}^1 d(\cos\theta)\cos\theta f_{kk'}^s\notag\\
&=f_1^s
\end{align}
Without loss of generality, we see that $\widetilde{f}_\ell^s=f_\ell^s$ for all $\ell\not=0$. 

We now move onto the derivation of the effective mass of quasiparticles in the Landau-Majorana liquid. The energy functional of the  Landau-Majorana liquid is given by

\begin{align}
\delta E[\delta n]&=\sum_{k\sigma} \widetilde{\epsilon}_k \delta n_{k\sigma}+\frac{1}{2V}\sum_{kk'\sigma\sigma'}\widetilde{f}_{\sigma\sigma'}(k,\,k')\delta n_{k\sigma}\delta n_{k'\sigma'}
\end{align}
where we have used the picture of $\widetilde{\epsilon}_{k}$ rather than the $\delta \widetilde{n}_{k\sigma}$ picture. With the above, we then see that
\begin{align}
\frac{\delta \widetilde{E}}{\delta n_{k\sigma}}=\widetilde{\epsilon}_k+\frac{1}{V}\sum_{k\sigma'}\widetilde{f}_{\sigma\sigma'}\delta n_{k'\sigma}
\end{align}

\noindent where we have foregone the spin index in the quasiparticle energy. We now expand the relevant quantities, noting that ${\bf q}\cdot \hat{k}=q$ and ${\bf q}\cdot \hat{k}'=q\cos\theta$:
\begin{align}
\widetilde{\epsilon}_{k-q}&\approx\widetilde{\epsilon}_k-{\bf q}\cdot \nabla_k \widetilde{\epsilon}_k\notag\\
&\approx \epsilon_k f-q v_M^* 
\end{align}
\begin{align}
n_{k'-q}^0&\approx n_{k'}^0-{\bf q}\cdot \nabla \widetilde{\epsilon}_{k'}\frac{\partial n_{k'}^0}{\partial \epsilon_{k'}}\notag\\
&\approx n_{k'}^0-qv_M^* \cos\theta \frac{\partial n_{k'}^0}{\partial \widetilde{\epsilon}_{k'}}
\end{align}

\noindent We therefore see that
\begin{align}
\frac{\delta \widetilde{E}}{\delta n_{k\sigma}}&\approx \epsilon_k f-qv_M^* +\frac{1}{V}\sum_{k\sigma'}\widetilde{f}_{\sigma \sigma'}(k,\,k')\left\{
-qv_M^* \cos\theta \frac{\partial n_{k'}^0}{\partial \widetilde{\epsilon_{k'}}}
\right\}
\end{align}
Now, note that the current $j_k=k_F/m$ is given by
\begin{align}
j_k&=-\frac{\partial}{\partial q}\left(\frac{\delta E}{\delta n_{k-q}}\right)\bigg|_{q=0}
\end{align}
Hence,
\begin{align}
j_k&=v_M^*+\frac{1}{V}\sum_{k\sigma}\widetilde{f}_{\sigma\sigma'}(k,\,k')v_M^*\frac{\partial n_{k'}^0}{\partial \epsilon_{k}'}\cos\theta\notag\\
&=v_M^* \left(1+\frac{\widetilde{F}_1^s}{3}\right)\notag\\
&=\frac{v_F^*}{2}\left(1+\frac{k_F}{4\alpha}\right)\left(1+\frac{F_1^s}{3}\right)
\end{align}
where we have used the fact that $\widetilde{F}_1^s=F_1^s$. Therefore, the Landau-Majorana liquid effective mass $\widetilde{m}^*$ is given by
\begin{align}
\frac{\widetilde{m}^*}{m}&=\frac{1}{2}\left(1+\frac{k_F}{4\alpha}\right)\left(1+\frac{F_1^s}{3}\right)\notag\\
&\approx \frac{k_F}{8\alpha}\left(1+\frac{F_1^s}{3}\right)\notag\\
&=\frac{k_F}{8\alpha}\frac{m^*}{m}
\end{align}
The ratio of the Landau-Majorana liquid effective mass and the Landau-Fermi liquid effective mass is then given by
\begin{align}
\frac{\widetilde{m}^*}{m^*}\approx \frac{k_F}{8\alpha}
\end{align}
This subsequently leads to the expression in the text.
\section{Appendix B: Expressions for the Landau-Majorana zero sound and the Pomeranchuck instability condition in the density}
The equation describing the zero sound renormalized velocity $s$ when $s>>1$ is given by Eq. \eqref{main}. The solutions $s_1$, $s_2$, and $s_3$ are given below: 
{\small 
\begin{subequations}
\begin{equation}
s_1=\frac{A_0}{3\widetilde{v}}\left\{
F_0^s+\frac{G^{(1)}\sqrt[3]{2}}{ \sqrt[3]{G^{(2)}+\sqrt{{F_0^s}^3\{(G^{(2)})^2-4(G^{(1)})^3\}}}}+\frac{\sqrt[3]{G^{(2)}+\sqrt{{F_0^s}^3\{(G^{(2)})^2-4(G^{(1)})^3\}}}}{F_0^s\sqrt[3]{2}}
\right\}
\end{equation}

\begin{equation}
s_2=\frac{A_0}{3\widetilde{v}}\left\{
F_0^s-\frac{(1+i\sqrt{3})G^{(1)}}{2^{2/3}\sqrt[3]{G^{(2)}+\sqrt{{F_0^s}^3\{(G^{(2)})^2-4(G^{(1)})^3\}}}}
-\frac{(1-i\sqrt{3})\sqrt[3]{G^{(2)}+\sqrt{{F_0^s}^3\{(G^{(2)})^2-4(G^{(1)})^3\}}}}{2^{4/3}F_0^s}
\right\}
\end{equation}

\begin{align}
s_3&=\frac{Aa_0}{3\widetilde{v}}\left\{
F_0^s-\frac{(1-i\sqrt{3})G^{(1)}}{2^{2/3}\sqrt[3]{G^{(2)}+\sqrt{{F_0^s}^3\{(G^{(2)})^2-4(G^{(1)})^3\}}}}
-\frac{(1+i\sqrt{3})\sqrt[3]{G^{(2)}+\sqrt{{F_0^s}^3\{(G^{(2)})^2-4(G^{(1)})^3\}}}}{2^{4/3}F_0^s}
\right\}\notag\\
&=s_2^*
\end{align}

\end{subequations}
}
where we define

\begin{subequations}
\begin{equation}
G^{(1)}(\widetilde{v},\,F_0^s)\equiv G^{(1)}=({F_0^s})^3+\widetilde{v}^2\left\{({F_0^s})^2+2F_0^s+1\right\}
\end{equation}
\begin{equation}
G^{(2)}(\widetilde{v},\,F_0^s)\equiv G^{(2)}=2{F_0^s}^6+3(F_0^s \widetilde{v})^2\left\{({F_0^s})^3 +5({F_0^s})^2+7{F_0^s}+3\right\}
\end{equation}
\end{subequations}

By assuming a large magnitude $|F_0^s|$ of the interaction, we can solve for the Pomeranchuck instability condition by solving for $\widetilde{v}$ when
\begin{align}
&\phantom{+}4 \left(9 (F_0^s \widetilde{v}+\widetilde{v}) \left({F_0^s}^2 (-\widetilde{v})-F_0^s \widetilde{v}\right)-9 {F_0^s}^4\right)^3\notag\\
&+\left(54 {F_0^s}^6+81
   {F_0^s}^5 \widetilde{v}^2+405 {F_0^s}^4 \widetilde{v}^2+567 {F_0^s}^3 \widetilde{v}^2+243 {F_0^s}^2 \widetilde{v}^2\right)^2=0
\end{align}
which will lead to a vanishing imaginary portion of $s_2$ and $s_3$. Excluding the trivial case of $\widetilde{v}=0$, we find that
{\small 
\begin{align}
\widetilde{v}&=\frac{1}{2} \sqrt{\frac{3}{2}} \sqrt{\frac{-{F_0^s}^7+14 {F_0^s}^6+93 {F_0^s}^5+212 {F_0^s}^4+233
   {F_0^s}^3+126 {F_0^s}^2\pm \sqrt{{F_0^s}^2 ({F_0^s}+1)^9 ({F_0^s}+9)^3}+27 {F_0^s}}{({F_0^s}+1)^6}}\\
&\approx \begin{cases}
&\frac{1}{2} \sqrt{21-\frac{3 \left({F_0^s}^7+\sqrt{{F_0^s}^2 ({F_0^s}+1)^9 ({F_0^s}+9)^3}\right)}{2 {F_0^s}^6}}\approx 2\sqrt{3}
\notag\\
&\frac{1}{2}\sqrt{\frac{3}{2}}\sqrt{-2{F_0^s}}=\frac{\sqrt{3}}{2}\sqrt{-{F_0^s}}
\end{cases}
\end{align}
}
Invoking the form of $\widetilde{v}$ defined previously, we can then readily solve for the expressions given in the text.
\vspace{20mm}
\bibliographystyle{iopart-num}
\bibliography{main}{}

\end{document}